\documentclass{rspublic}

\usepackage[dvips]{graphicx}
\usepackage{epsfig,epsf}
\usepackage{natbib}
\usepackage{sublabel}
\usepackage{latexsym}
\usepackage{amsthm}
\usepackage{amssymb}
\usepackage{graphicx}
\usepackage{amsmath}
\usepackage{pstricks}
\usepackage{fullpage}
\newpsobject{showgrid}{psgrid}{subgriddiv=1,griddots=10,gridlabels=6pt}

\newcommand {\norm} [1] { \lVert #1 \rVert}

\newcommand {\abs} [1] {\left| #1 \right|}

\numberwithin{equation}{section}

\begin{document}
\title[Pipe flow at high $Re$]{The critical layer in pipe flow
at high Reynolds number}

\author[D. Viswanath]{D.Viswanath $^{1}$}
\affiliation{$^1$ 
Mathematics Department, University of Michigan, Ann Arbor, MI 48109.
}
\label{firstpage}

\maketitle

\begin{abstract}{pipe flow, traveling waves, critical layer, GMRES-hookstep}
We report the computation of a family of traveling wave solutions of
pipe flow up to $Re=75000$. As in all lower-branch solutions, streaks
and rolls feature prominently in these solutions. For large $Re$,
these solutions develop a critical layer away from the wall. Although
the solutions are linearly unstable, the two unstable eigenvalues
approach $0$ as $Re\rightarrow\infty$ at rates given by $Re^{-0.41}$
and $Re^{-0.87}$ --- surprisingly, the solutions become more stable as
the flow becomes less viscous.  The formation of the critical layer
and other aspects of the $Re\rightarrow\infty$ limit could be
universal to lower-branch solutions of shear flows.  We give
implementation details of the GMRES-hookstep and Arnoldi iterations
used for computing these solutions and their spectra, while pointing
out the new aspects of our method. 
\end{abstract}

\section{Introduction}

In this article, we look at a lower-branch traveling wave solution in
the $Re\rightarrow\infty$ limit. The traveling wave we chose to
compute has an asymmetric arrangement of streaks, with two fast
streaks located preferentially on one side of the pipe. \cite{SEY}
found that states with such an asymmetry arise in direct numerical
simulations of transition to turbulence. \cite{PK} computed such a
traveling wave using a bifurcation point of a mirror-symmetric family
around $Re=1000$. Our computations of the same traveling wave go up to
$Re=75000$ and help elucidate aspects of the $Re\rightarrow \infty$
asymptotic limit. 

\begin{figure}
\begin{center}
(a)
\includegraphics[height=2.2in,width=2in]{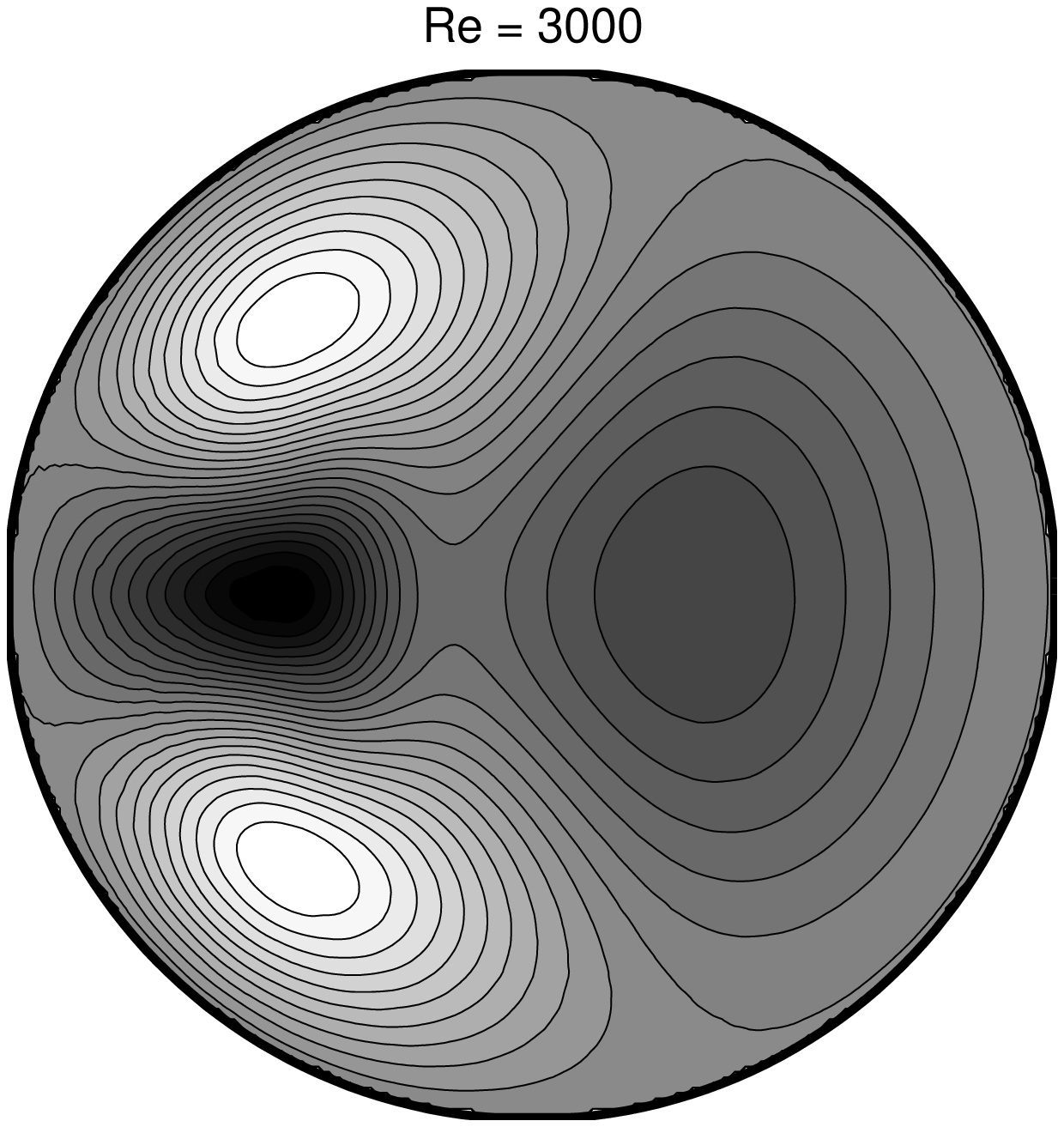}
\hspace*{.3cm}
(b)
\includegraphics[height=2.2in,width=2in]{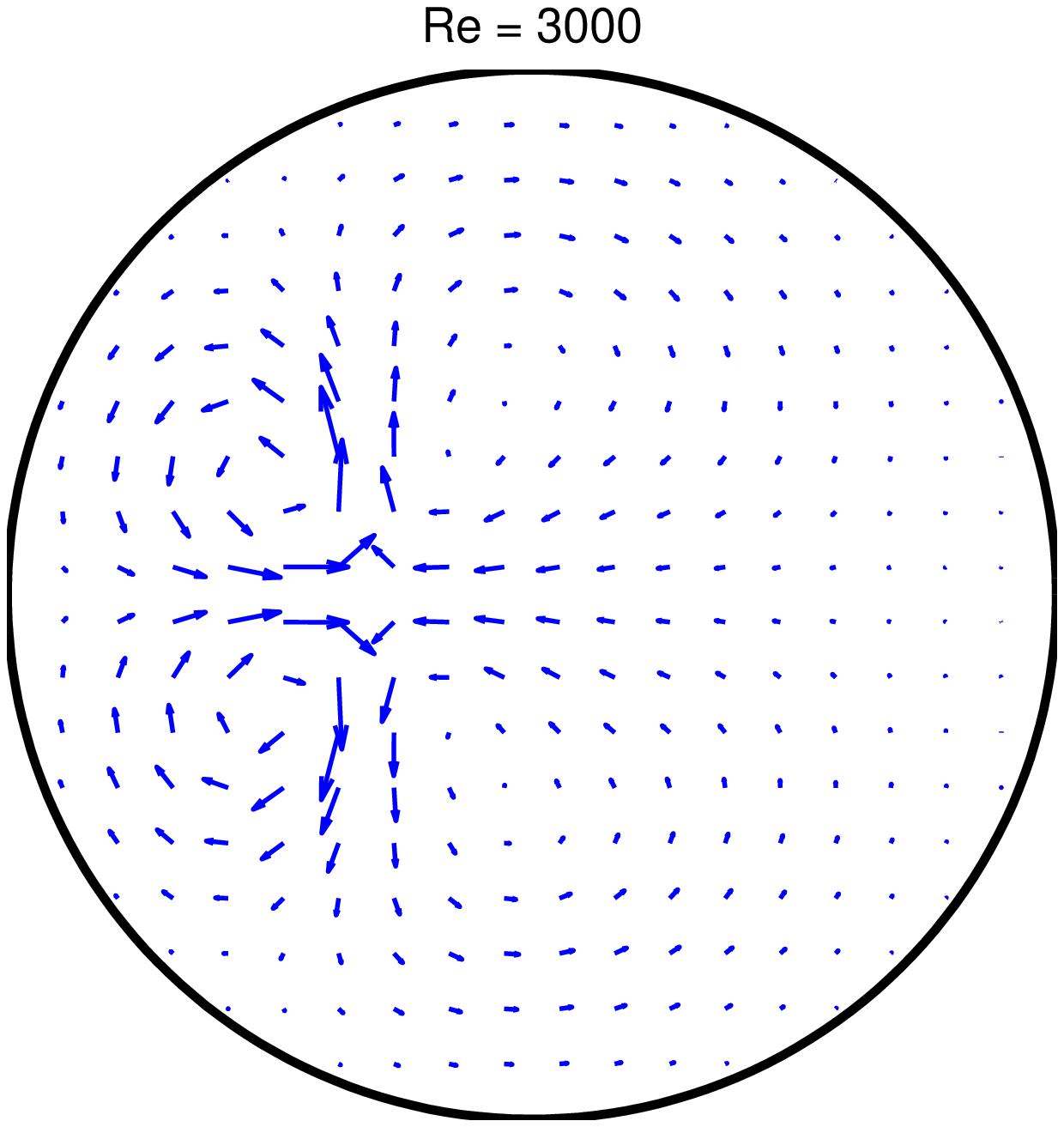}
\end{center}
\caption[xyz]{(a): Contour plot of the $z$-averaged streamwise
velocity with the laminar flow subtracted. The contour levels are 
equispaced in $(-0.18, 0.16)$, with the red (or lighter) regions on the left
of the pipe being the high-speed streaks. 
(b): The rolls are shown using
a quiver plot of the $z$-averaged radial and azimuthal velocities.
The maximum magnitude of a velocity vector in the quiver plot is
$.0065$.
}
\label{fig-1}
\end{figure}

The fast streaks near the wall are the most prominent and stable
structures in lower-branch traveling wave solutions of pipe flow
\citep{FE2, WK}. The fast streaks are regions in a circular
section where the streamwise velocity significantly exceeds the
laminar value. The fast and slow streaks can form different patterns.
The pattern that characterizes some of the computed solutions is an
invariance with respect to rotation about the pipe axis by $2\pi/m$,
where $m=2,3,4,\ldots$. The rolls, which correspond to positive and
negative streamwise vorticity, form complementary patterns.  Although
the computed solutions use periodic boundary condition in the axial
direction and very short pipes, they do pick structures that
transitional pipe flow tends to develop \citep{HD, WillisK2}.  The
data analysis techniques used to extract these patterns are set up to
pick patterns with rotational symmetry \citep{ESHW, SEV, WillisK2}.
The streak pattern of the asymmetric traveling wave does not have any
$m$-fold rotational symmetry as evident from Figure \ref{fig-1}.

 \cite{WGW} (also see \citep{Waleffe3}) showed that the $Re\rightarrow\infty$
limit of a symmetric lower-branch solution of plane Couette flow is
characterized by a number of features. The streaks remain $O(1)$, but
the magnitude of the rolls and of the fundamental and higher
streamwise modes decrease algebraically with $Re$. The scaling exponents
for the rolls and the fundamental streamwise mode of the asymmetric traveling
wave are $-1.08$ and $-0.97$, which may be compared with $-1$ and
$-0.9$ for the symmetric solution of plane Couette flow.
Higher streamwise modes decrease even faster.


 The most important consequence of these
scalings is the development of a critical layer away from the circular
boundary of the pipe. The theory of \cite{WGW} successfully identifies
the critical curve as given by $w_0(r,\theta) = c_z$, where
$w_0$ is the $z$-averaged streamwise velocity and $c_z$ is
the wavespeed in the $z$ direction. The fundamental component
of the radial velocity is concentrated in a region around the critical
curve and drops off to zero away from that region. We find that the
size of the region decreases at the rate $Re^{-0.32}$ as $Re$
increases, which compares well with the rate of $Re^{-1/3}$ derived by
\cite{WGW} using formal arguments.  The exponents for the rates at
which the sizes of the regions decrease with $Re$ are different for the
fundamental mode of the streamwise velocity and the mean streamwise
vorticity. These are found to be $-0.26$ and $-0.23$, respectively, in
Section 4. These exponents present a
challenge to asymptotic theory.


At the end of Section 4, we suggest that it might be useful to 
calculate the analogue of the critical curve for puffs. Puffs have
a well-defined extent and travel down the pipe with a well-defined
speed. The analogue of the critical curve would be a surface,
embedded inside the puff, on all points of which the streamwise
velocity equals the speed of the puff. Such a surface could be
helpful in elucidating the structure of the puff.

The Newton equations for solving a nonlinear system can sometimes be
solved efficiently in a Krylov subspace \citep{BS, SNGS}.  We point
out two new aspects of the extensions to the Newton-Krylov procedure
introduced by
\cite{Viswanath1}.  The first novelty is the formulation of the Newton
equations. In the case of pipe flow, the formulation allows for
translation of the velocity field along the pipe axis or rotation
around the pipe axis. 
The second novelty is the GMRES-hookstep combination explained
in Section 5.


For large $Re$, the lower-branch asymmetric traveling wave looks very
different from both the laminar solution of pipe flow and the sort of
turbulence that is typically observed at such $Re$. Unlike the laminar
solution, the traveling wave develops streaks, for instance.  Unlike
fully developed turbulence, there is no rapid decay of correlations.
The form of the asymmetric traveling wave is nearly independent of the
$z$ direction at high $Re$. Thus one may ask if the lower-branch
solutions are relevant for high $Re$ turbulence and if they can be
realized in the lab. The answer to the first question is probably no.
The second question is a difficult challenge to experiment.  That the
computations are restricted to small pipes is less of an issue for
high $Re$ because of the scaling of the streamwise modes mentioned
above and discussed in Section 3.


\section{Preliminary data}

\begin{table}
\begin{center}
\begin{tabular}{c|c|c|c|c|c|c|c|c|c}
$Re$ & $L$ & $M$ & $N$ & $T$ & $c_z$ & $I,D$ & $KE$   & $\lambda_1$
& $\lambda_2$\\\hline
1500  &  81 & 18 & 16 & 10 &$.7339$  &1.1051 &0.9772    &0.0463 &0.0149\\\hline
10000 & 101 & 24 & 16 & 10 &$.8236$  &1.0657 &0.9781    &0.0189 &0.0022\\\hline
75000 & 151 & 24 & 4  & 15 &$.8715$  &1.0460 &0.9829    &         & 
\end{tabular}
\end{center}
\caption[]{The column headings are explained in the text. The eigenvalues
$\lambda_i$ were not computed at $Re=75000$. }
\label{table-1}
\end{table}

The asymmetric traveling waves were computed at a number of values of
$Re$ in the range $1500\leq Re\leq 75000$. Some basic data is
summarized in Table \ref{table-1}.  The choice of units and boundary
conditions follows that of
\cite{FE1}. The pipe radius is chosen as the unit of length. The
unit of velocity is equal to the centerline velocity of the
Hagen-Poiseuille laminar flow. The Reynolds number is based on the
pipe radius, centerline velocity of the Hagen-Poiseuille laminar flow,
and the kinematic viscosity. The boundary condition is no-slip at the
pipe wall and periodic in the axial direction.  The mass-flux of the
flow, which is fixed at $0.5$, drives the flow.  The pipe length or
period is $2\pi\Lambda$. We took $\Lambda = 1/1.44$, but this choice
has no special significance in the $Re\rightarrow\infty$ limit.

The quantities $L, M, N$ listed in Table \ref{table-1} parameterize
the spatial grid used to represent the velocity field. The spatial
coordinate system $r,\theta, z$ was cylindrical, with $u,v,w$ being
the three components of velocity, respectively.  The three components
of vorticity are denoted as $\xi,\eta,\zeta$. The radial component of
the velocity field ${\bf u}$ is represented as
\begin{equation}
u(r,\theta, z) = \sum_{\substack{-M< m < M\\-N<n<N}}
\hat{u}_{n,m}(r) \exp(i m \theta) \exp(i n z/\Lambda),
\label{eqn-2-1}
\end{equation}
with the discretization using $2M$ and $2N$ Fourier points along
$\theta$ and $z$, respectively. The coefficients $\hat{u}_{n,m}(r)$
are even in $r$ for $m$ odd, and odd for $m$ even. Each
$\hat{u}_{n,m}$ is represented using its values at the Chebyshev
points $r=\cos(\pi i/L)$, $0\leq i\leq (L-1)/2$. Note that $L$ is
always odd. The vorticity component $\xi$ has an analogous
representation.  As the velocity field has zero divergence, the entire
velocity field can be recovered using $u$, $\xi$, $\bar{v}$, and
$\bar{w}$, where $\bar{v}(r)$ and $\bar{w}(r)$ are averages of $v$ and
$w$ with respect to both $\theta$ and $z$. After setting the modes
with $\abs{m}=M$ or $\abs{n}=N$ to zero, we are left with
$(L-2)+((2N-1)(2M-1)-1)(L-3)/2$ independent degrees of freedom.

In terms of the modes, the boundary conditions become
$\hat{u}_{n,m}(1) = \hat{\xi}_{n,m}(1) = 0$ and
$\frac{\partial\hat{u}_{n,m}(1)}{\partial r} = 0$.  The constant mass
flux condition implies a pressure gradient along $z$ that can change
from instant to instant for an evolving flow.

The wavespeed of the traveling wave is given by $c_z$.  To find each
traveling wave, one solves for a velocity field ${\bf u_0}$ such that
${\bf u}(r,\theta,z,t) = {\bf u_0}(r,\theta,z-c_z t)$ is a solution of
the Navier-Stokes equation. The artificial parameter $T$, which occurs
in Table \ref{table-1}, arises in the solution procedure and its
meaning is explained in Section 5.

The rate of energy dissipation per unit mass is given by
$2 D/Re$, where $D$ is the integral of 
\begin{equation*}
\frac{1}{4 \pi^2\Lambda}
\biggl(
\frac{1}{r^2}(u^2+v^2 - 2 u_{\theta} v + 2u v_{\theta}) + 
\sum_{U=u,v,w} \biggl(\frac{\partial U}{\partial r}\biggr)^2
+  \biggl(\frac{\partial U}{\partial z}\biggr)^2
+  \frac{1}{r^2}\biggl(\frac{\partial U}{\partial \theta}\biggr)^2
\biggr)
\end{equation*}
over the volume of the pipe. The rate of energy input per unit mass is given
by $2 I/Re$, where 
\begin{equation*}
I = -\frac{Re}{4\pi^2\Lambda}\int \nabla\cdot(p{\bf u}),
\end{equation*}
with $p$ being pressure and with the integral being over the volume
of the pipe. The friction coefficient \citep{WK} is the same as $I$, but
with a different normalization. $D$ and $I$ are normalized to be $1$ for
the laminar flow.  From Table \ref{table-1}, we see that $D=I$ for
all the traveling waves in agreement with energy conservation.
Kinetic energy, denoted KE in Table \ref{table-1}, is also normalized
to evaluate to $1$ for laminar flow. 

The Navier-Stokes equation for pipe flow, with periodic boundary along
$z$, is unchanged by the shift-reflect transformation. The 
shift-reflect transformation reflects the velocity field about the plane
$\theta = 0$ or $\theta = \pi$, and shifts it along $z$ by half a pipe
length. All the asymmetric traveling waves have only two unstable
eigenvalues in the shift-reflection symmetric subspace. Those are
given as $\lambda_1$ and $\lambda_2$ in Table \ref{table-1}. Section 6
has a discussion of the spectrum of the traveling waves.

\section{Scaling of modal kinetic energies}
\begin{figure}
\begin{center}
(a)\includegraphics[height=2.3in,width=2.3in]{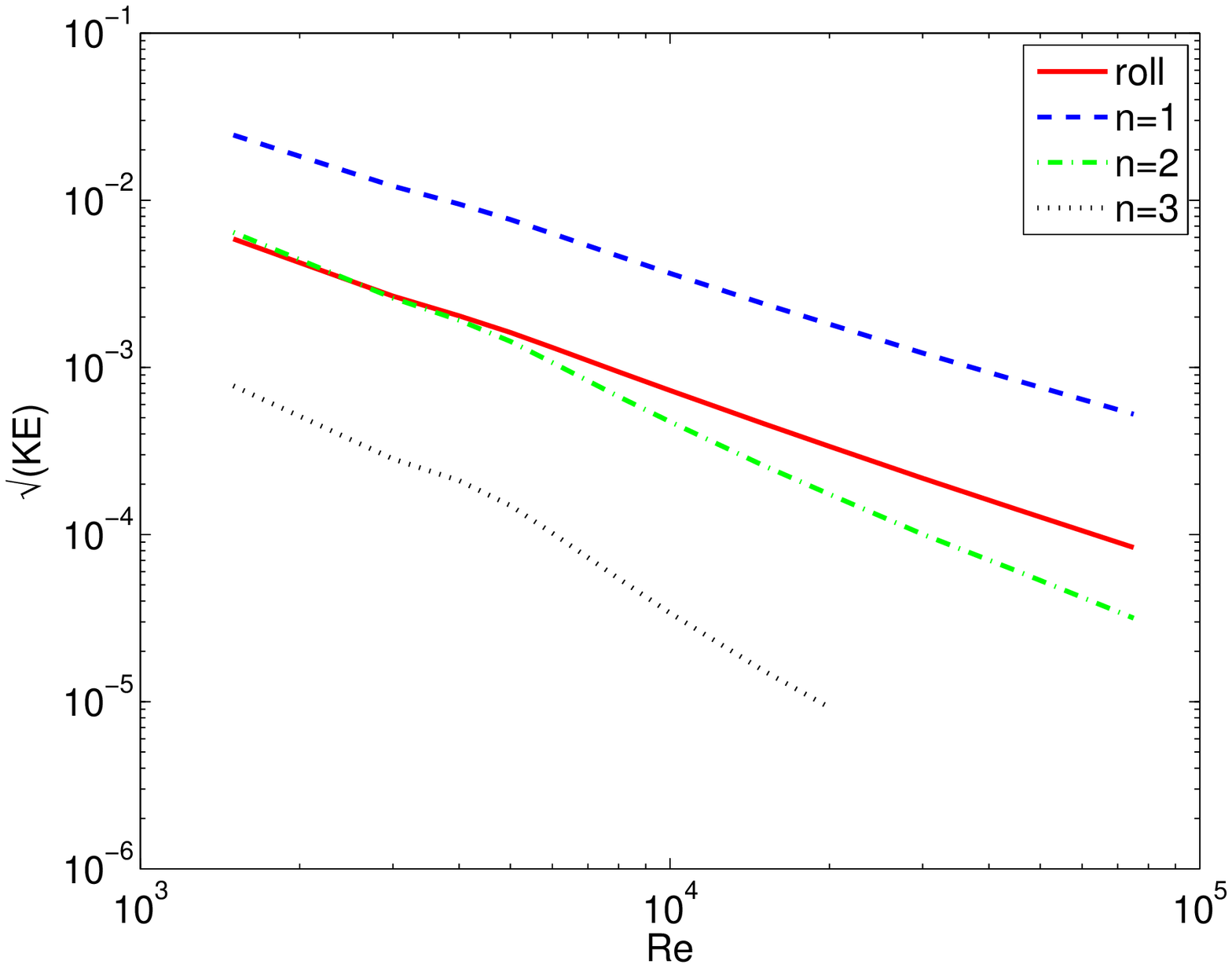}
\hspace*{0.3cm}
(b)\includegraphics[height=2.3in,width=2.3in]{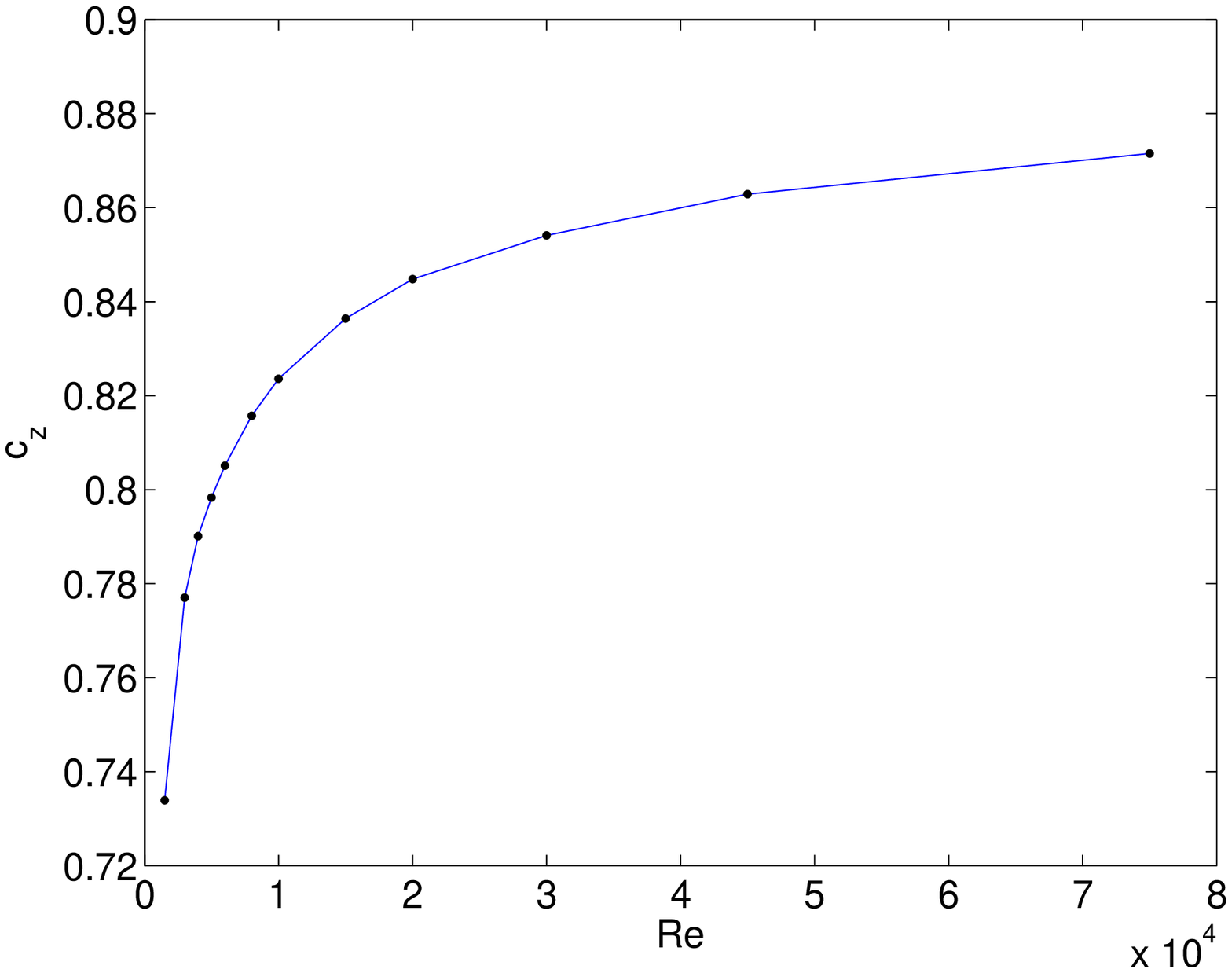}
\end{center}
\caption[xyz]{(a): The magnitude of a mode is measured using the square root
of the kinetic energy (KE). The index $n$ is used to pick modes from
Fourier expansions of the form \eqref{eqn-2-1}. 
(b): The dependence of wavespeed on $Re$.}
\label{fig-2}
\end{figure}

Figure \ref{fig-2}a shows the variation of the kinetic energy in
various modes as a function of $Re$. To find the kinetic energy for
the $n=1$ streamwise mode, we form Fourier expansions of type
\eqref{eqn-2-1} for $v$ and $w$ as well.  The volume integral for
kinetic energy is computed by setting all modes with $ n \neq \pm 1$
equal to zero.  The kinetic energies of the other streamwise modes are
computed in a similar manner.

The kinetic energy of the rolls  is obtained using $n=0$ mode only,
but the $w$ component is set to zero.  Retaining
only the $n=0$ modes is equivalent to averaging the velocity field
with respect to $z$.  The $z$-averaged $w$ corresponds to streaks.

As evident from Figure \ref{fig-2}a, the magnitudes of the modes
decrease with $Re$ algebraically and are proportional to $Re^{e}$ for
high $Re$ and a suitable exponent $e$. The exponents for the rolls,
$n=1$, $n=2$, and $n=3$ obtained using $Re\geq 8000$ were $-1.08$,
$-0.97$, $-1.35$, and $-1.92$, respectively. For comparison, the
exponents for rolls and the $n=1$ mode are $-1$ and $-0.9$ for the symmetric
lower-branch solution of plane Couette flow \citep{WGW}.

\begin{figure}
\begin{center}
(a)
\includegraphics[height=2.2in,width=2in]{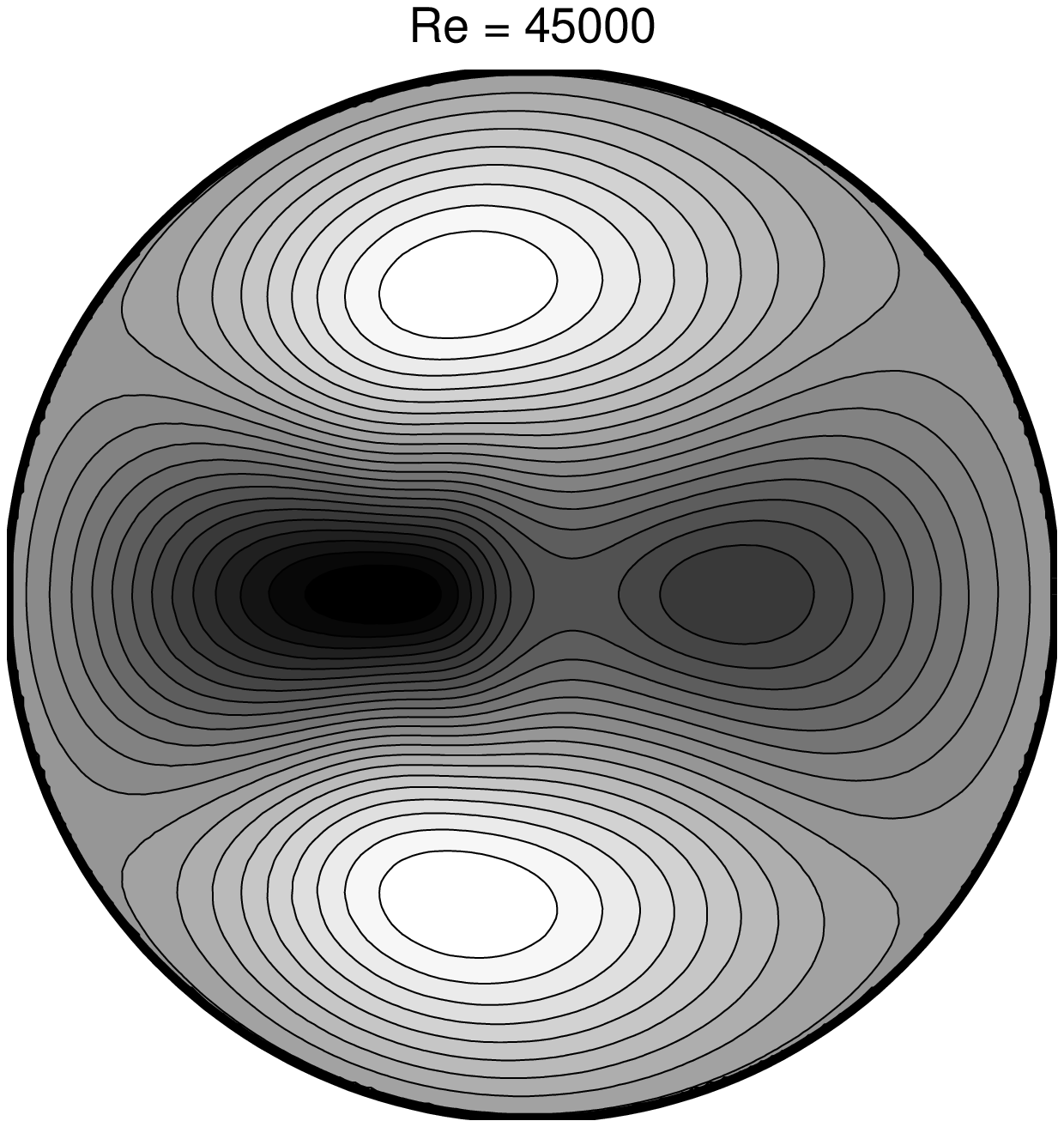}
\hspace*{.3cm}
(b)
\includegraphics[height=2.2in,width=2in]{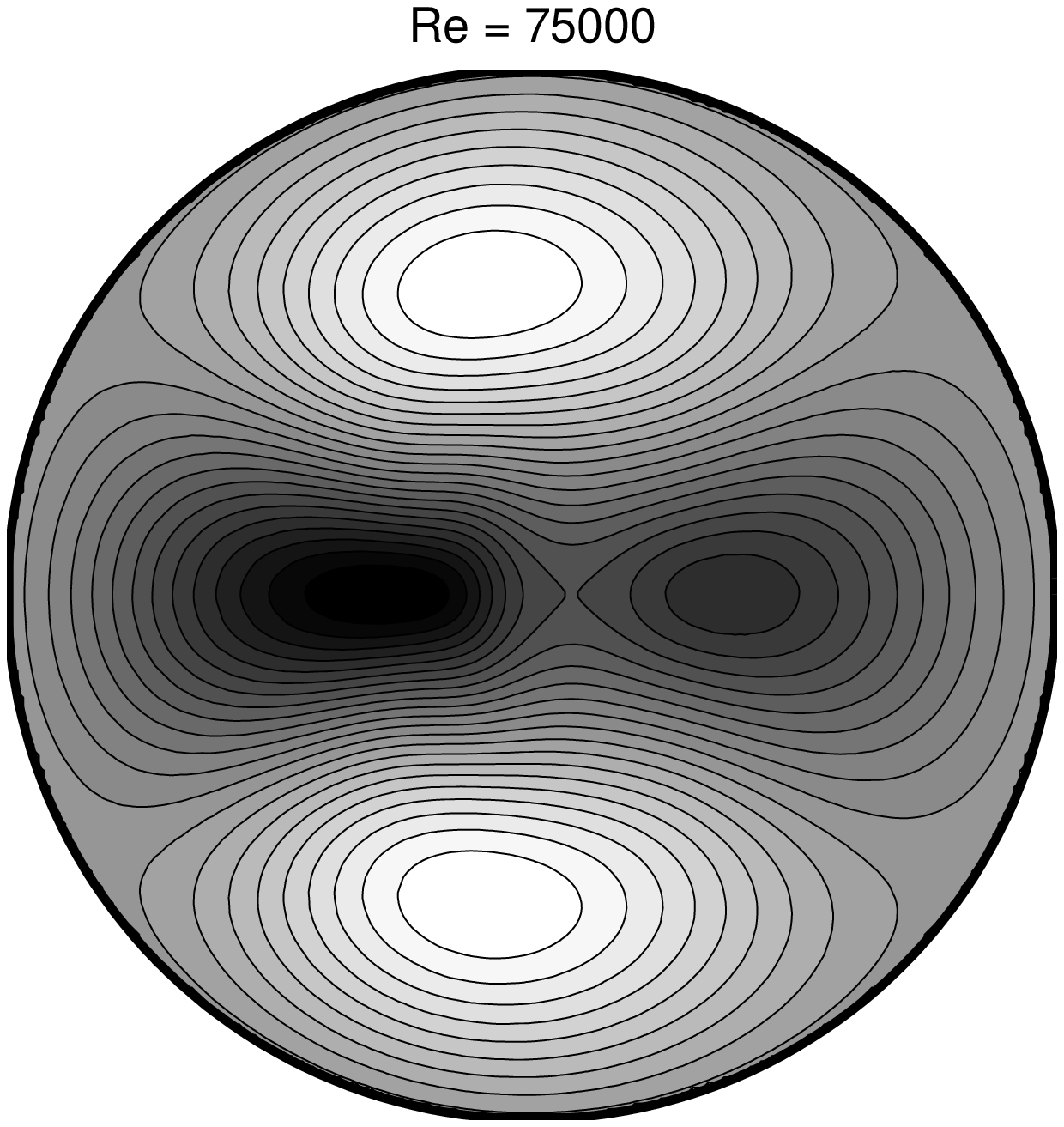}
\end{center}
\caption[xyz]{The plots of streaks are as in Figure \ref{fig-1}.
The contours in (a) and (b) are equispaced in $(-0.11, 0.16)$ and
$(-0.11,0.15)$, respectively.  
}
\label{fig-3}
\end{figure}

Figure \ref{fig-3}b shows that the wavespeed $c_z$ increases with
$Re$. An application of Wynn's $\rho$-algorithm \citep{Wynn}
shows the limit of
$c_z$ as $Re\rightarrow\infty$ to be $0.88$. The speed of the
asymmetric traveling wave is nearly twice the speed of 
puffs in transitional pipe flow. In our units, the speed of the
puff is about $0.45$ around $Re=2000$ \citep{PM}.

From Figure \ref{fig-3}, we conclude that the streaks converge as
$Re\rightarrow\infty$ and that the plots in that figure are a good
approximation to the limit. Those plots differ quite a bit from the
plot at $Re=3000$ in Figure \ref{fig-2}, with the position of the two
high-speed streaks being much more to the left of the pipe at
$Re=3000$.

\section{The critical layer}

The Fourier expansion of $u$ \eqref{eqn-2-1} can be rewritten as $$u =
u_0(r,\theta) + u_1(r,\theta) \exp(iz/\Lambda) +
u^{\ast}_1(r,\theta)\exp(-iz/\Lambda)+\cdots,$$
where the asterisk denotes
complex conjugation.  Similar expansions can be formed for $v$, $w$,
and the vorticity components. To illustrate the critical layer, 
we will begin by looking at $\abs{u_1}$.

\begin{figure}
\begin{center}
(a)
\includegraphics[height=1.8in,width=1.7in]{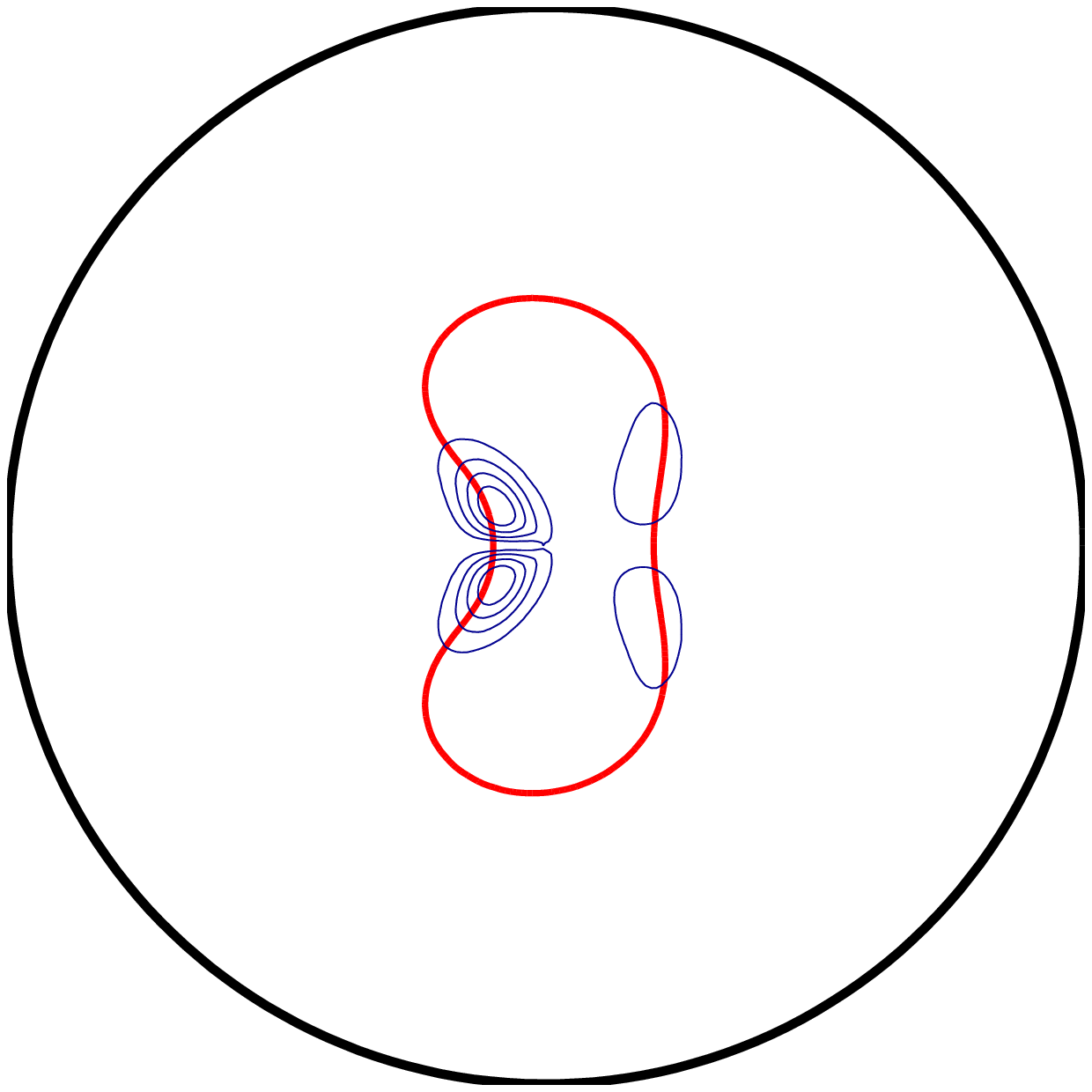}
(b)
\includegraphics[height=1.8in,width=1.7in]{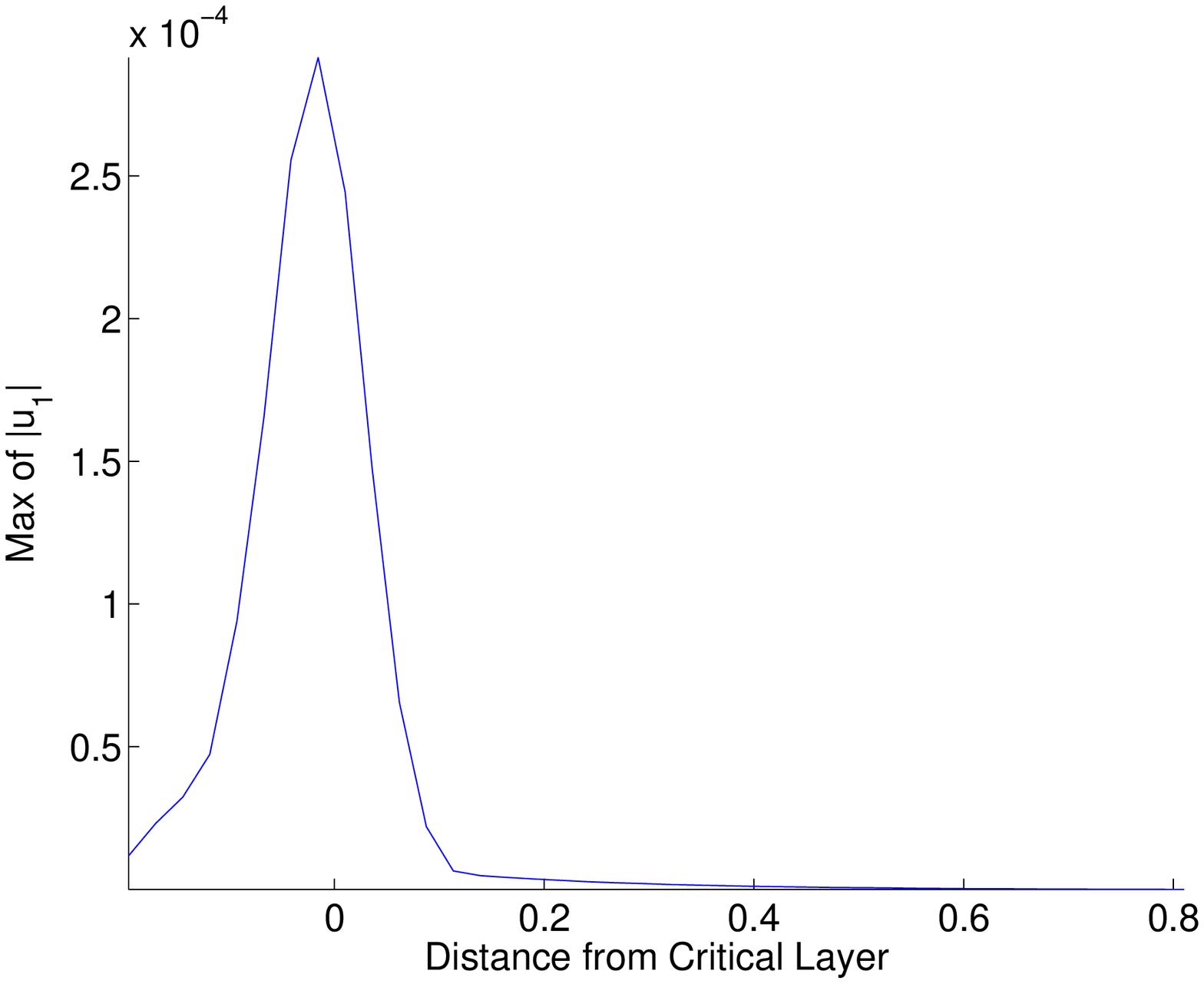}
(c)
\includegraphics[height=1.9in,width=1.7in]{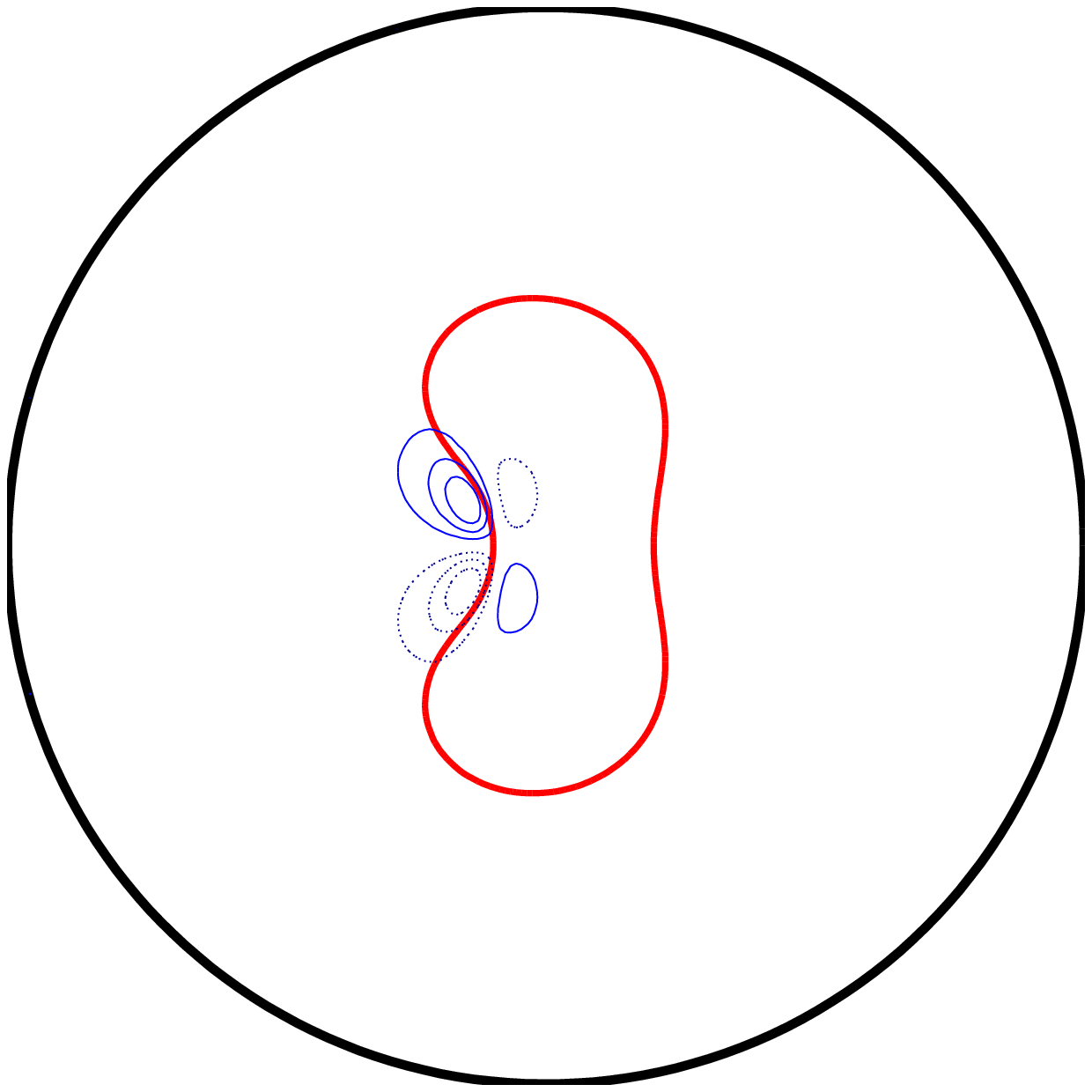}
\end{center}
\caption[xyz]{All plots at $Re=75000$. (a): The red and thick curve is
the critical curve $w_0(r,\theta)=c_z$. The four values for contouring
$\abs{u_1}$ were equispaced between $0$ and $\max\abs{u_1}$. 
(b): The maximum of $\abs{u_1}$ is
taken over curves all points of which are at the
distance $d$ from the critical curve. The distance $d$, which
is the $x$-axis of the plot, is negative inside the
critical curve and positive outside.
(c): Contour plots of $z$-averaged streamwise vorticity $\zeta_0$.
The solid and dashed lines correspond to positive and negative
$\zeta_0$.
}
\label{fig-4}
\end{figure}

\begin{figure}
\begin{center}
\includegraphics[height=2in,width=2in]{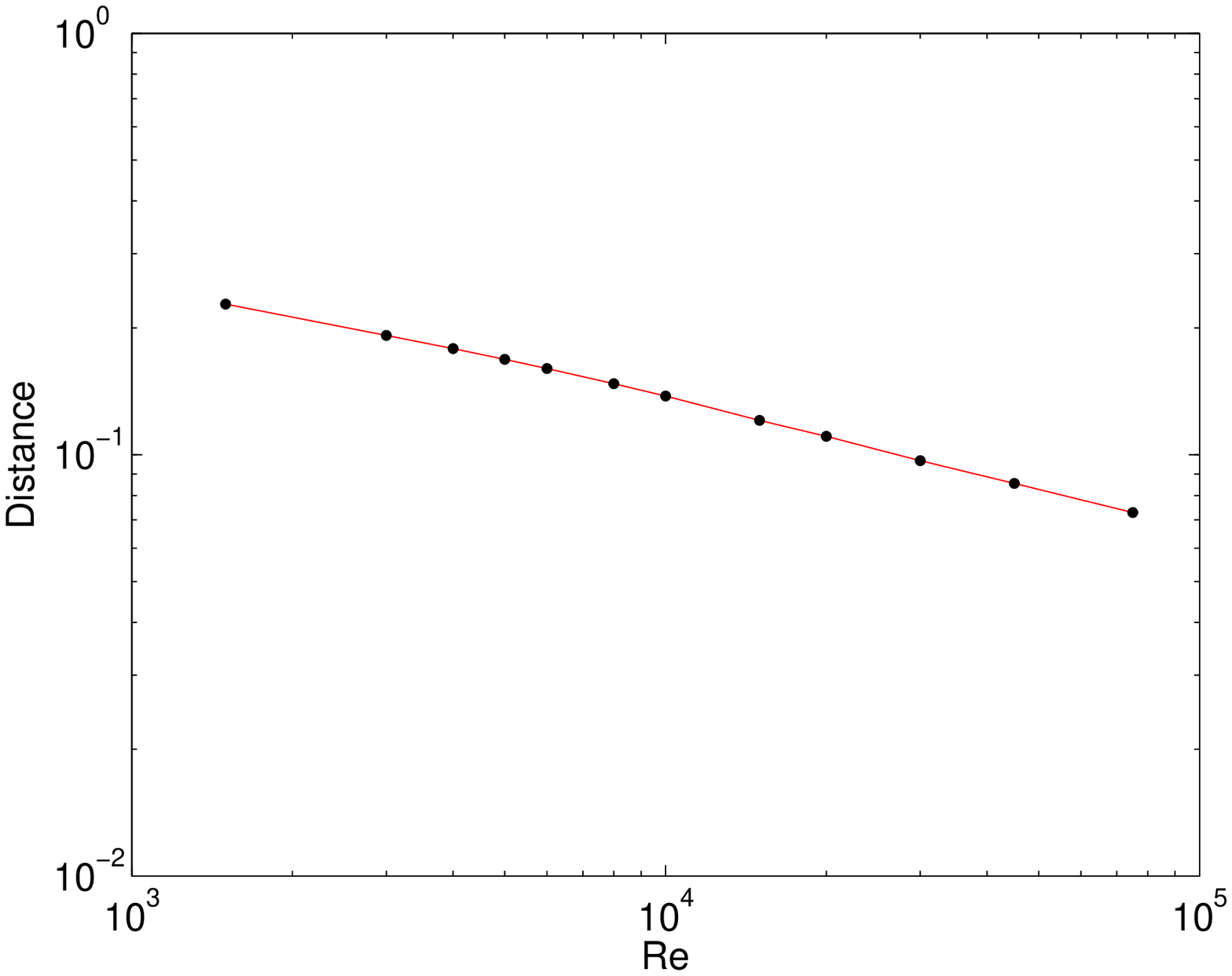}
\includegraphics[height=2in,width=2in]{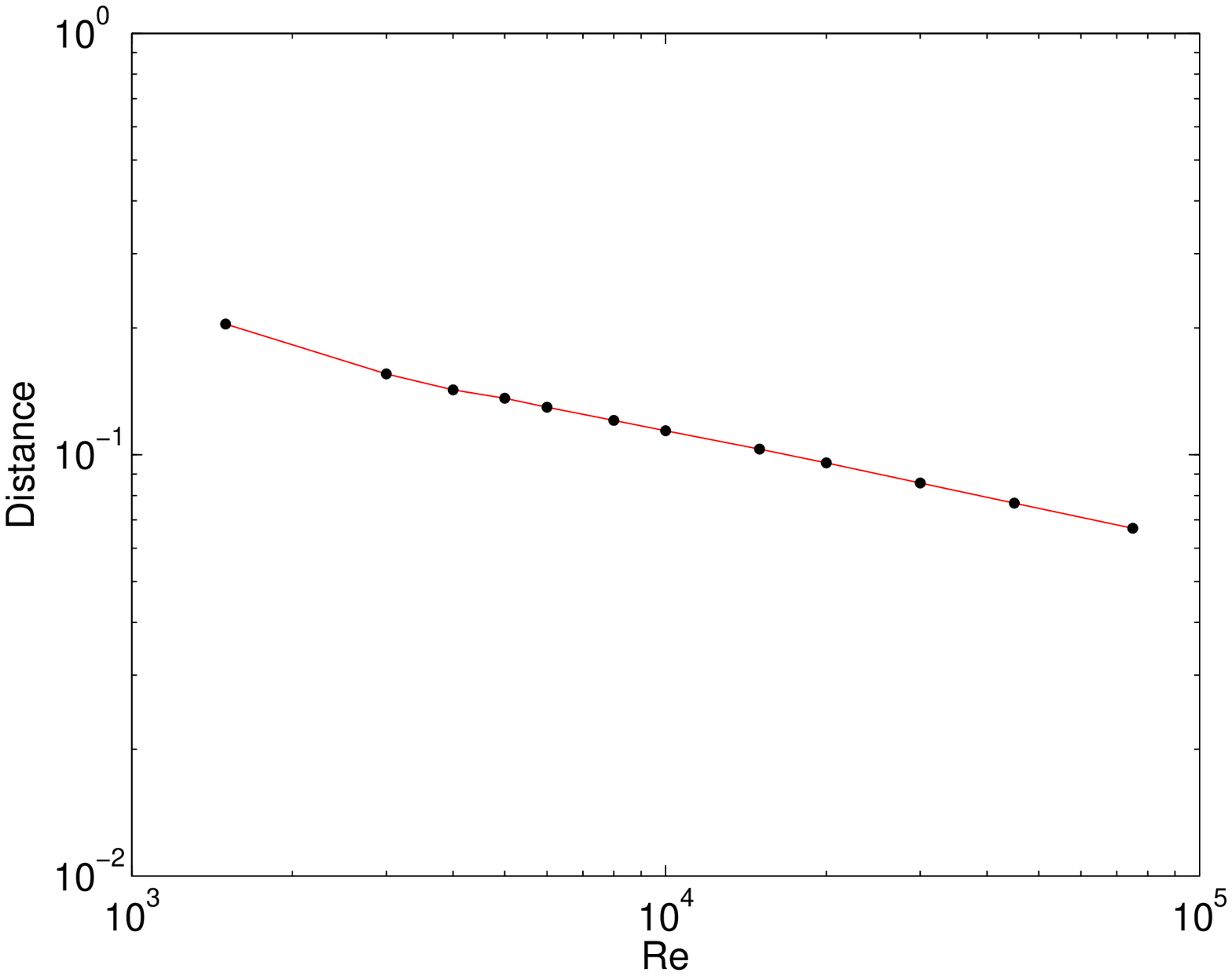}
\includegraphics[height=2in,width=2in]{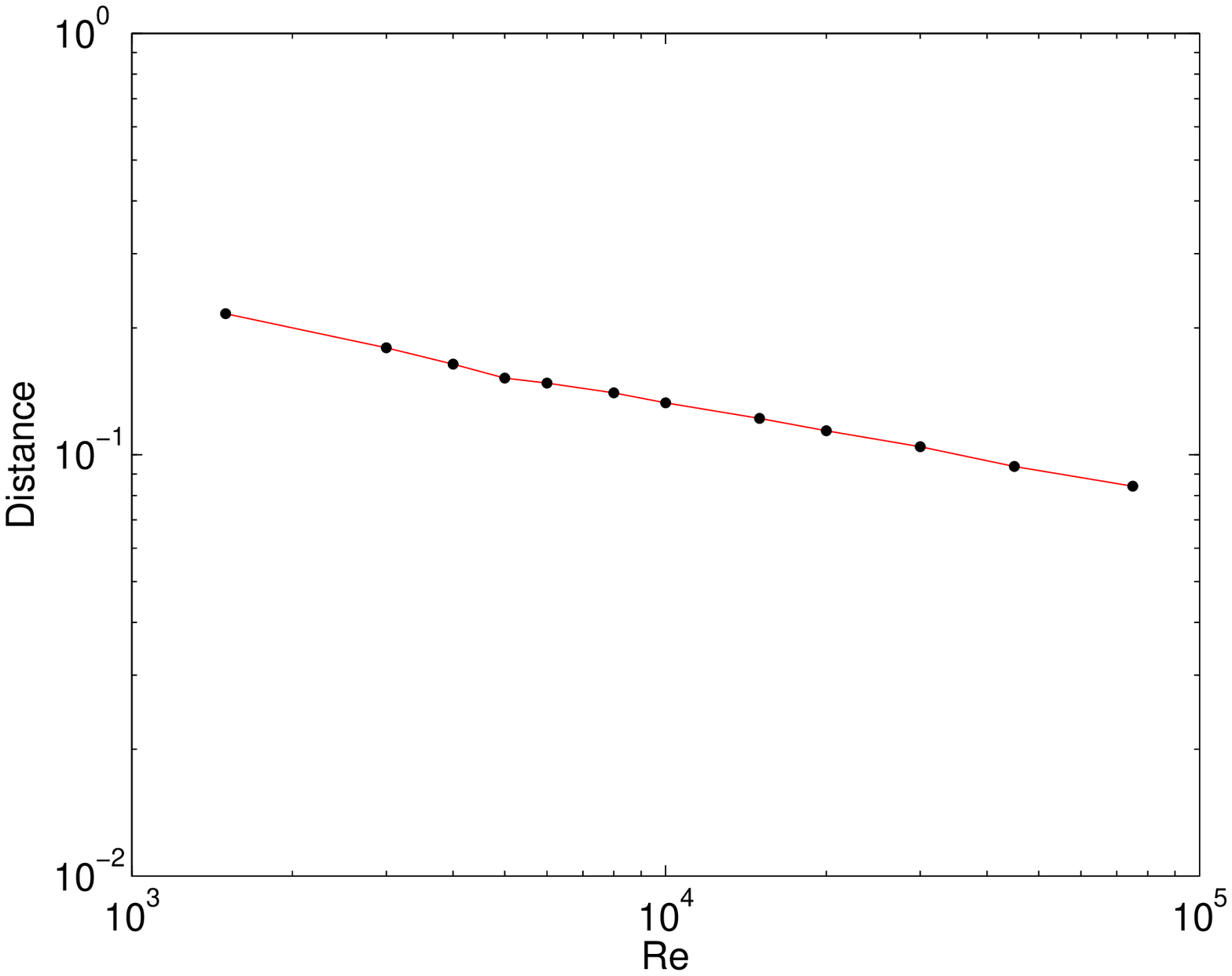}
\end{center}
\caption[xyz]{The plots correspond to $\abs{u1}$, $\abs{w1}$,
and $\zeta_0$, respectively. The plots show that the width of
the critical layer decreases with $Re$ at different rates for
$\abs{u_1}$, $\abs{w_1}$, and $\zeta_0$.}
\label{fig-6}
\end{figure}

\cite{WGW} derived the equation $w_0(r,\theta) = c_z$ 
for the critical curve.
The critical curve is shown as a thick red curve in Figure
\ref{fig-4}a.  It is closer to the center of the pipe than to the pipe
wall.  The contour lines of $\abs{u_1}$ are all nestled around the
critical curve. In particular, the contour lines occur as two groups
near the indentation at the left of the critical curve. This compares
well with Figure 3 of \cite{WGW}. Figure \ref{fig-4}b shows that
$\abs{u_1}$ takes its maximum value on or very close to the critical
curve and falls off rapidly away from the critical curve. The first
two plots of Figure
\ref{fig-4} give a good idea of how $\abs{u_1}$ varies inside the
unit circle. The critical region is a band around the critical curve
where most of the variation of $\abs{u_1}$ and certain other
quantities is concentrated. The band need not be of uniform width.


Figure \ref{fig-4}c shows contour plots of $\zeta_0$. The regions
where $\zeta_0$ is positive or negative agree very well with the
position of the rolls. Counter-rotating vortices are a well-known
feature of lower-branch solutions and of small perturbations of the
laminar flow that trigger turbulence. Like rolls and streamwise modes,
the scaling of whose magnitudes with $Re$ is shown in Figure
\ref{fig-2}, the magnitude of $\zeta_0$ also decreases with $Re$.



From Figure \ref{fig-4}c, it is evident that most of the
variation of $\abs{u_1}$ is in a region around the critical curve.
Similar plots can be produced for $\abs{w_1}$ or $\zeta_0$.
In such plots the peaks become noticeably sharper as $Re$
increases.

The purpose of Figure \ref{fig-6} is to estimate the rate at which the
contour curves, such as those in Figure \ref{fig-4}a and c, approach
the critical curve as $Re\rightarrow\infty$.  For each value of $Re$,
a specific contour curve is picked. For $\abs{u_1}$, $\abs{w_1}$, and
$\zeta_0$, the chosen contour curve is for half their maximums. We
pick the point on the contour curve that is farthest from the critical
curve and plot its distance against $Re$.  Such plots are a good way
to measure the thickness of the critical region. They follow the
convention where the width of a Gaussian density function is measured
at half its maximum.

Fits using $Re\geq 8000$ show that the thickness scales as
$Re^{-0.32}$, $Re^{-0.26}$, and $Re^{-0.23}$ for $\abs{u_1}$,
$\abs{w_1}$, and $\zeta_0$, respectively.  The exponents do not change
appreciably if fits are made by dropping the data points with
smaller $Re$.


Perhaps the main achievement of \cite{WGW} is to give a formula for
the critical curve.  In the context of pipe flow, the critical curve
is the set of all points $(r,\theta)$ such that
$w_0(r,\theta)=c_z$. We have used that formula throughout this
section. Their calculations apply directly to $\abs{u_1}$ and
$\abs{v_1}$, and predict that the contour curves of those quantities
will approach the critical curve at a rate given by $Re^{-1/3}$. The
exponent that we found for $\abs{u_1}$, which came in at $-0.32$, is
in excellent agreement with that prediction.  The exponents for
$\abs{w_1}$ and $\zeta_0$ indicate that the contour curves of those
quantities concentrate more slowly on the critical curve than those of
$\abs{u_1}$. A more refined theory is probably needed to explain those
exponents.

The thickness of the critical layer is highly unlikely to be uniform
around the critical curve. The manner in which the thickness varies
along the critical curve appears worthy of investigation. It appears
that the variation of the thickness could be related to the structure
of the rolls.  Even at low $Re$, such as $Re=1500$, contour plots
still show that structures tend to develop around the critical
curve. This motivates a suggestion that will end this section.

Puffs are structures observed in transitional pipe flow that have
a well-defined extent. They travel down the pipe with a well-defined
speed. It could be interesting to calculate the surface formed by
all points of the puff whose streamwise velocity equals the speed at
which the puff moves down the pipe. Such a surface would be the
analogue of the critical curve for a lower-branch traveling wave.

\section{Implementation of GMRES-hookstep and Arnoldi iterations}

In section 2, we pointed out that the velocity field for pipe flow
with suitable boundary conditions can be recovered from $\bar{v}$,
$\bar{w}$, $u$ and $\xi$. If we pack the information in those
variables into a single column vector $x$ with real components, it is
possible to recover the entire velocity field given $x$. $X(t;x)$ is
the column vector that results from allowing the flow to evolve for
time $t$.  To compute $X(t; x)$, a velocity field is constructed
starting from $x$ and then allowed to evolve for time $t$ using a
direct numerical simulation code. $X(t;x)$ is constructed from the
final velocity field. We have generally used Runge-Kutta methods with
constant step sizes (except for the last step) to compute $X(t;x)$.
The reason is that the discretized flow is then a dynamical system 
that is smooth and close to the Navier-Stokes flow. Adaptive time 
stepping strategies introduce non-smoothness and imply that the
discretized flow is no longer a dynamical system.

The methods for computing traveling waves and other solutions that
will be described depend upon the shear flow mainly in the computation
of $X(t;x)$. The other dependence is in the definition of the
translation operators. Given the Fourier representation
\eqref{eqn-2-1} of $u(r,\theta,z)$, the representation after a
translation along the axis and a rotation about the axis is given
by
\begin{equation}
u(r,\theta+s_\theta, z + s_z) 
=
\sum_{\substack{-M< m < M\\-N<n<N}}
\hat{u}_{n,m}(r) 
\exp(ims_\theta)
\exp(ins_z/\Lambda)
\exp(i m \theta+i n z/\Lambda).
\label{eqn-5-1}
\end{equation}
We use linear operators defined by
\begin{align}
\mathcal{T}_1 u(r,\theta, z) &=
\sum_{\substack{-M< m < M\\-N<n<N}} i m \hat{u}_{n,m}(r)\exp(im\theta)\exp(inz/\Lambda)\nonumber\\
\mathcal{T}_2 u(r,\theta, z) &=
\sum_{\substack{-M< m < M\\-N<n<N}} (in/\Lambda)  \hat{u}_{n,m}(r)\exp(im\theta)\exp(inz/\Lambda)
\label{eqn-5-2}
\end{align}
to effect the translation and the rotation in \eqref{eqn-5-1}. In particular,
$$u(r,\theta+s_\theta, z+s_z) = \exp(s_\theta \mathcal{T}_1)
\exp(s_z\mathcal{T}_2)u(r,\theta,z).$$ 
The definition of the linear operators
$\mathcal{T}_i$ depends upon the shear flow. The definition 
of the linear operators for plane Couette flow is identical to
that for pipe Poiseuille flow \citep{Viswanath1}. The operators
$\mathcal{T}_i$ can be made to act on a vector $x$ that encodes
a velocity field in an obvious way, by making them act on
each component of the velocity field. Then
$\exp(s_\theta \mathcal{T}_1)
\exp(s_z\mathcal{T}_2)x$ encodes a translated and rotated 
velocity field. Expressing the translation and rotation of 
a velocity field using $\mathcal{T}_i$ makes it possible to
differentiate with respect to $s_\theta$ and $s_z$ while
deriving the Newton equations.

Given the ability to compute $X(x;t)$ and the linear operators
of \eqref{eqn-5-2}, the numerical methods described in this section
need to know nothing more about the shear flow. Determining the exact
dimension of the vector $x$ can be a little tricky because one needs
to eliminate Fourier coefficients that are conjugates of certain others
and so on \citep{Viswanath1}. It is unlikely that one may leave out 
some essential components as this error will become manifest when
trying to construct the velocity field from $x$. It is more likely
that $x$ may end up having duplicates. In principle, that would
make some of the matrices that occur later singular. In practice,
the effect of having duplicates in $x$ will probably introduce some
error without being disastrous. 

A big part of the numerical method for computing traveling waves,
relative periodic orbits, and other solutions that will now be
described are the well-known GMRES and Arnoldi
iterations. \cite{TrefethenBau} give a lucid account of these methods
and more importantly their convergence properties. Pointers to
the original literature can be found in the end notes of their book
or in many other well-known textbooks of numerical linear algebra.

\subsection{GMRES-hookstep iteration}

A relative periodic orbit is a solution of the Navier-Stokes equation
where the initial velocity field evolves for time $T$, which is the 
period, to reach a certain final state. In the case of pipe flow,
it must be possible to translate the final velocity field along the
axis and then rotate it to get back the initial velocity. 
If $x_0$ encodes the initial velocity field,
\begin{equation}
\exp(-s_\theta \mathcal{T}_1)
\exp(-s_z\mathcal{T}_2) X(T; x_0) = x_0,
\label{eqn-5-3}
\end{equation}
where $s_\theta$ and $s_z$ are shifts in the azimuthal and streamwise
directions, respectively. To find a relative periodic orbit, one
must solve for $x_0$, $s_\theta$, $s_z$, and the period $T$ such
that the nonlinear equation \eqref{eqn-5-3} is satisfied.

A relative periodic orbit is the most general object that our method
can find. Periodic orbits are a special case where $s_\theta = s_z =
0$. Traveling waves are a special case where $T$ is fixed to be a
small but not too small number. A traveling wave will satisfy
\eqref{eqn-5-3} for any $T>0$ and suitably chosen $s_\theta, s_z$. But
there is no guarantee that $x_0$ merely translates and rotates as it
evolves. In other words, the solution of \eqref{eqn-5-3} could be a
relative periodic orbit that is not a traveling wave. $T$ is chosen
small enough to make it likely that the solution of \eqref{eqn-5-3} is
a traveling wave, although it is not important to have a small $T$ if
we already know that the initial guess for $x_0$ is near a traveling
wave. An equilibrium or steady solution can also be thought of as
a special case of a relative periodic orbit. The reason for treating
traveling waves as special cases of relative periodic orbits is
explained at the end of this section.


Suppose $\tilde{x}_0, s_x, s_z, T$ is an initial guess to a solution
of \eqref{eqn-5-3} and that 
\begin{equation}
y_0 = \exp(-s_\theta \mathcal{T}_1)
\exp(-s_z\mathcal{T}_2) X(T; \tilde{x}_0).
\label{eqn-5-4}
\end{equation}
Linearizing one gets the
following Newton equations \citep{Viswanath1}:
\begin{equation}
\begin{pmatrix}
\exp(-s_\theta \mathcal{T}_1)
\exp(-s_z\mathcal{T}_2) 
\frac{\partial X(T; \tilde{x}_0)}{\partial\tilde{x}_0} - I &
-\mathcal{T}_1 y_0 & 
-\mathcal{T}_2 y_0 & f(y_0) \\
\mathrm{Transpose}(\mathcal{T}_1 \tilde{x}_0) & 0 & 0 & 0 \\
\mathrm{Transpose}(\mathcal{T}_2 \tilde{x}_0) & 0 & 0 & 0 \\
\mathrm{Transpose}(f(\tilde{x}_0)) & 0 & 0 & 0 
\end{pmatrix}
\begin{pmatrix}
\delta x\\
\delta s_\theta\\
\delta s_z\\
\delta T
\end{pmatrix}
=
\begin{pmatrix}
\tilde{x}_0 - y_0\\
0\\
0\\
0
\end{pmatrix}.
\label{eqn-5-5}
\end{equation}
In the above system, $I$ is the identity matrix whose dimension
equals that of $\tilde{x}_0$; and $f(x)$ is such that  $dx/dt = f(x)$
is the spatially discretized Navier-Stokes equation written in terms of
the vector $x$ which encodes the discretized velocity field. The code
for evaluating $f(x)$ can be extracted from a direct numerical simulation
code with a little work. One can also approximate $f(x)$ as 
$(X(h;x) - x)/h$, where $h$ is small. We have not tried approximating
$f(x)$ using differences, but it is probably fine to do so. 
The last three rows of the linear system \eqref{eqn-5-5} correspond
to phase conditions \citep{Viswanath1}.

To find a relative periodic orbit, one step of the Newton iteration
would be to solve \eqref{eqn-5-5} for the $\delta$s and add those
corrections to the initial guess. To find a traveling wave,
\eqref{eqn-5-5} must be modified by dropping the last row and the last
column because $T$ is fixed. If the traveling wave has the
shift-reflect symmetry, as the traveling wave family studied in this
paper does, then $s_\theta = 0$, because rotation around the
pipe axis breaks that symmetry. In such a case, we must drop the first
and the third of the last three columns, and likewise with the
rows. To find an equilibrium solution, we must drop the last three
columns and rows. All the special cases of a relative periodic orbit
mentioned above can be dealt with in this manner. In each case, we
denote the resulting linear system as $A\Delta = b$.

To solve such a linear system using a Krylov subspace method like GMRES,
it is not necessary to invert $A$ nor is it even necessary to form $A$
explicitly. It is enough if $A$ can be applied to vectors. The only
difficulty in applying $A$ to a vector arises in calculating 
$$
\exp(-s_\theta \mathcal{T}_1)
\exp(-s_z\mathcal{T}_2) 
\frac{\partial X(T; \tilde{x}_0)}{\partial\tilde{x}_0} c,$$
where
$c$ is a column vector of the same dimension as $\tilde{x}_0$. That quantity
can be calculated using differences as
\begin{equation}
\frac{\exp(-s_\theta \mathcal{T}_1)
\exp(-s_z\mathcal{T}_2)
X(T; \tilde{x}_0 + \epsilon c)-y_0}{\epsilon},
\label{eqn-5-6}
\end{equation}
where $\epsilon$ is chosen such that $\norm{\epsilon c} 
\approx 10^{-7} \norm{\tilde{x}_0}$. The choice of the norm will
be discussed shortly. Even when $\tilde{x}_0$ is nearly equal to 
$y_0$, which is defined by \eqref{eqn-5-4}, it is important not to
substitute $\tilde{x}_0$ for $y_0$ in \eqref{eqn-5-6}. 

The GMRES iteration for solving $A\Delta = b$ finds an
orthonormal matrix $Q_k$ at the $k$th stage such
that $A Q_{k} = Q_{k+1} H_{k+1,k}$ \citep{TrefethenBau}.
In implementing this step, it may be best to use the square root of
the kinetic energy of the vector field that $x$ encodes as the norm
over $x$.
At the $k$th stage GMRES would solve the least-squares
problem $\min_{y}\norm{H_{k+1,k} y -
\norm{b}e_1}$, where $y\in R^k$ and $e_1$ is the $k+1$ dimensional
vector with a $1$ at the top followed by $0$s. The approximation to
$\Delta$ at that stage would be $\Delta_k = Q_k y$. We do not attempt
to solve the Newton equation this way, however. The Newton equation is
useful only if the solution $\Delta$ is tiny enough that the
linearization that led to the Newton equation is valid.  That is often
not the case because the initial guesses are typically not so
accurate. A well-known solution is to minimize $\norm{A\Delta_\delta -
b}$ subject to the constraint $\norm{\Delta} \leq \delta$, where
$\delta$ has to be chosen small enough that the linearization within
that radius is valid \citep{DS}. The resulting step is called the
hookstep \citep{DS}.

We approximate the hookstep using GMRES as follows. To find
$\Delta_{\delta,k}$ that approximates the true hookstep
$\Delta_\delta$, we solve the minimization problem
\begin{equation}
\min_{y}\norm{H_{k+1,k} y -
\norm{b}e_1}
\label{eqn-5-7}
\end{equation}
subject to the constraint $\norm{y}\leq \delta$. That
minimization can be solved using the singular value decomposition
\citep{DS, GolubvanLoan}.
Let $H_{k+1,k} = U D V'$ be a reduced singular value 
decomposition ($V'$ is the transpose of the real unitary matrix $V$).
Let $p = (p_1,\ldots,p_k)'=\norm{b}U'e_1$.  If the diagonal entries
of the diagonal matrix $D$ are $d_i$, $q = (q_1,\ldots,q_k)'$ is found
using $q_i = p_i d_i/(\mu+d_i^2)$, $1\leq i\leq k$, where either $\mu>0$ is
such that $\norm{q} = \delta$ or $\mu=0$ if that
allows $\norm{q} \leq \delta$. Finding $\mu$ is an easy 1-dimensional
root finding problem. The solution of \eqref{eqn-5-7} is $y= V q$
and the GMRES-hookstep is $\Delta_{\delta,k} = Q_k y$. 

To complete the description of the GMRES-hookstep method, we have
to describe the choice of $k$, or the stopping criterion for 
finding a  $\Delta_{\delta,k}$ that approximates $\Delta_\delta$, and
also describe how $\delta$ is updated every time a new Newton system 
\eqref{eqn-5-5} is formed. There is a natural stopping criterion
for GMRES without the constraint $\norm{y}\leq \delta$. That is
because the relative residual error at the end of $k$ iterations can
be easily found as $r_k =
\norm{A\Delta_k-b}/ \norm{b}$.  For
GMRES-hookstep, we have no practical way of knowing how close
$\norm{A\Delta_{\delta,k}-b}$ is to $\norm{A\Delta_\delta - b}$.  Thus
there is no way to assess the quality of $\Delta_{\delta,k}$.  The
stopping criterion in our implementation is to pick a $k$ that is
large enough to ensure $r_k \leq .01$. In other words, we stop when
the GMRES iterate $\Delta_k$ is an acceptable substitute for the true
solution of $A\Delta = b$ believing then that the Krylov subspace
matrix $Q_k$ has enough column vectors to ensure that
$\Delta_{\delta,k}$ is an acceptable substitute for $\Delta_\delta$.
There is no theoretical support for this stopping criterion, but it
works very well in practice. 



The choice of $\delta$ follows standard trust-region prescriptions
\citep{DS}. The choice for $\delta$ for the very first GMRES-hookstep
iteration can be anything that looks reasonable. To assess the quality
of a $\delta$, we take $\norm{b} = \norm{\tilde{x}_0 - y_0}$ as the
error in the initial guess.  Once $\Delta_{\delta,k}$ is computed, we
update to $\tilde{x}_1 = \tilde{x}_0 +
\Delta_{\delta,k}(1:\mathrm{dim})$, where $\mathrm{dim}$ is the
dimension of $\tilde{x}_0$ and the subscripting of $\Delta_{\delta,k}$
follows MATLAB notation.  The quantities $s_\theta$, $s_z$, and $T$
are also updated, if applicable.  The linearization used to find
$\Delta_{\delta,k}$ predicts that the reduction in error in going from
$\tilde{x}_0$ to $\tilde{x}_1$ should be about $\norm{b} -
\norm{A\Delta_{\delta,k}-b}$. If the prediction is very good $\delta$
can be increased, and if it is bad $\delta$ must be decreased and a
new GMRES-hookstep must be computed. This completes the description of
the GMRES-hookstep method for solving \eqref{eqn-5-3}, each iteration
of which begins with a guess $\tilde{x}_0$ for $x_0$ and for the
shifts and the period, forms the Newton system \eqref{eqn-5-5}, uses
that Newton system to find $\Delta_{\delta,k}$, checks if $\delta$ is
acceptably small, and then uses $\Delta_{\delta,k}$ to form a better
guess. The iterations can be stopped if the error as measured by
$\norm{\tilde{x}_0 - y_0}/\norm{y_0}$ is less than the relative error
due to spatial discretization of the velocity field.


It is surprising that the method for computing $\Delta_{\delta,k}$ is
a new contribution considering it is quite a natural thing to do.
In an early paper on the use of Krylov subspaces for globally
convergent modifications of Newton's method, \cite{BS} formulated a
minimization problem ((4.2) of their paper) and called it the model
trust region problem. The solution to that problem is theoretically
equivalent to $\Delta_{\delta,k}$. The equivalence is similar
to that between GMRES and ORTHODIR, which predated GMRES, with our
formulation being more direct. We have described a practical method
for finding $\Delta_{\delta,k}$ with a criterion for choosing
$k$. We were not able to find implementations of GMRES-hookstep in the
literature, although one may exist that we were not able to track down.

Like the work of \cite{BS}, much of the later literature deals
with the dogleg and other strategies;
for instance see \citep{LV}.
The dogleg is an approximation
to the hookstep that is made up of only the gradient direction and the Newton
step \citep{DS}. It is preferred over the hookstep mainly because
its computation does not require the singular value decomposition.
Since the hookstep
moves away from the Newton step smoothly, one may suggest that
the Krylov subspace approximates the hookstep
bettr than the gradient. 
The dogleg is also much more complicated to implement within a 
Krylov subspace than the computation of $\Delta_{\delta,k}$
described here.  Having to compute the singular value decomposition
is not a problem because the way the Newton system
\eqref{eqn-5-5} is set up means that $k$ is small (being around $150$ at most
but more typically around $50$). 
Since the dogleg is only an approximation
to the hookstep, and is in fact harder to implement within a Krylov 
subspace, we see no reason to prefer it over the GMRES-hookstep
method.

\subsection{Arnoldi iteration}
Ignoring spatial discretization errors, the eigenvalues $\mu_i$ of the
matrix 
\begin{equation}
\exp(-s_\theta \mathcal{T}_1)
\exp(-s_z\mathcal{T}_2) 
\frac{\partial X(T; x_0)}{\partial x_0}
\label{eqn-5-8}
\end{equation}
are the eigenvalues of the corresponding relative periodic or periodic
solution. If $x_0$ encodes the velocity field of a traveling wave or
a relative periodic solution, then $\mu_i = \exp(\lambda_i T)$ where
$\lambda_i$ are the eigenvalues of the traveling wave or the equilibrium
solution.

The matrix \eqref{eqn-5-8} will be dense and large, but it can be
applied to vectors as in \eqref{eqn-5-6}. The Arnoldi iteration
forms $Q_k$, $Q_{k+1}$, and $H_{k+1,k}$ like GMRES, with the one difference
being that the starting vector $b$ is arbitrary. We usually take
$x_0$ as the starting vector but either rotate and translate it
or add some noise to ensure that it does not have the shift-reflect
symmetry. 
In the case of both pipe and channel flows, the laminar solution must be
subtracted from $x_0$ to get the right boundary conditions.
If $H_k$ is the matrix obtained by dropping the last row
of $H_{k+1,k}$, and $H_k y = \mu y$, then $\mu$ is an approximation
for an eigenvalue of \eqref{eqn-5-8} with $Q_k y$ being an approximation
for the corresponding eigenvector.

\begin{figure}
\begin{center}
(a)\includegraphics[height=1.5in,width=1.8in]{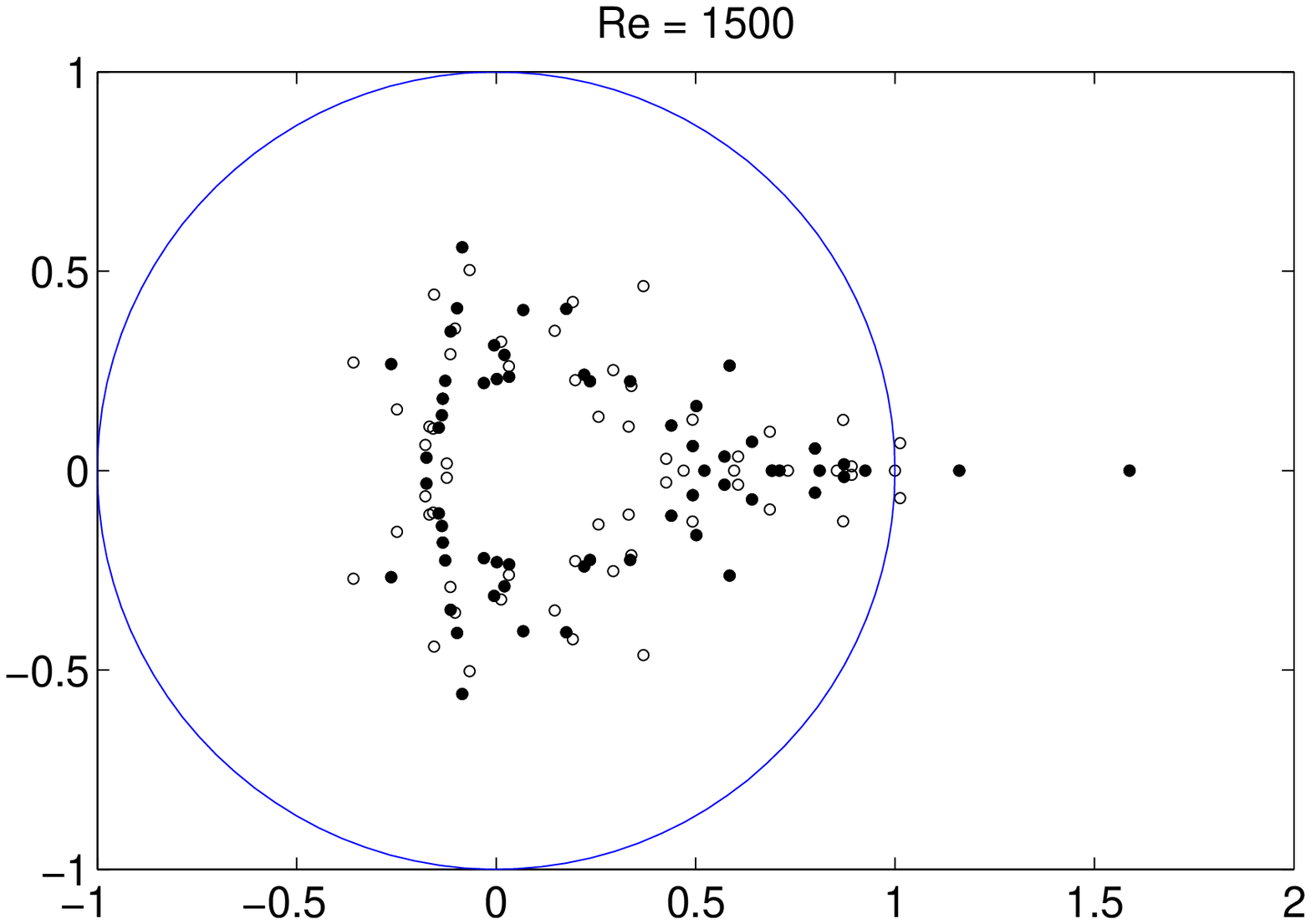}
(b)\includegraphics[height=1.5in,width=1.8in]{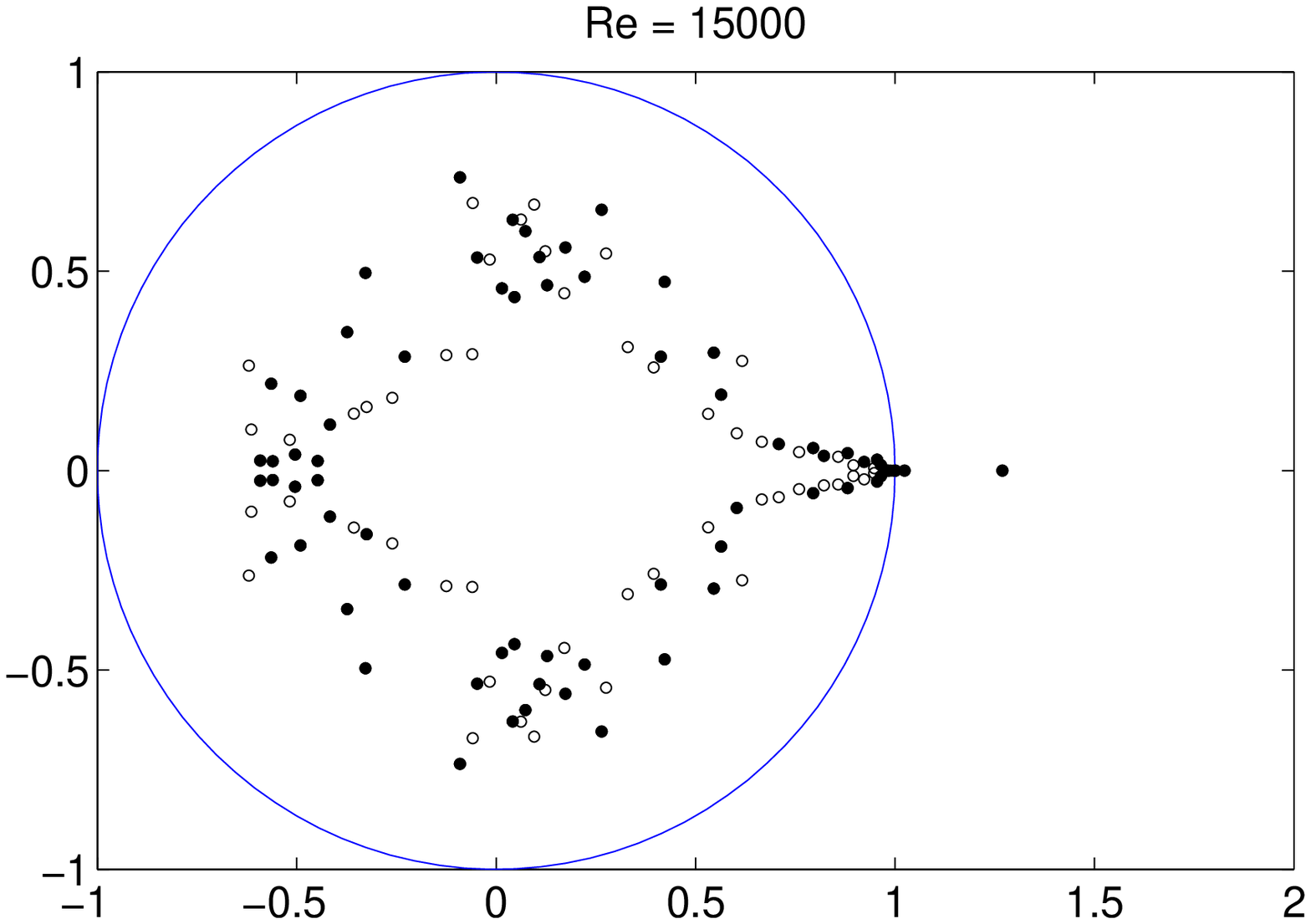}
(c)\includegraphics[height=1.5in,width=1.8in]{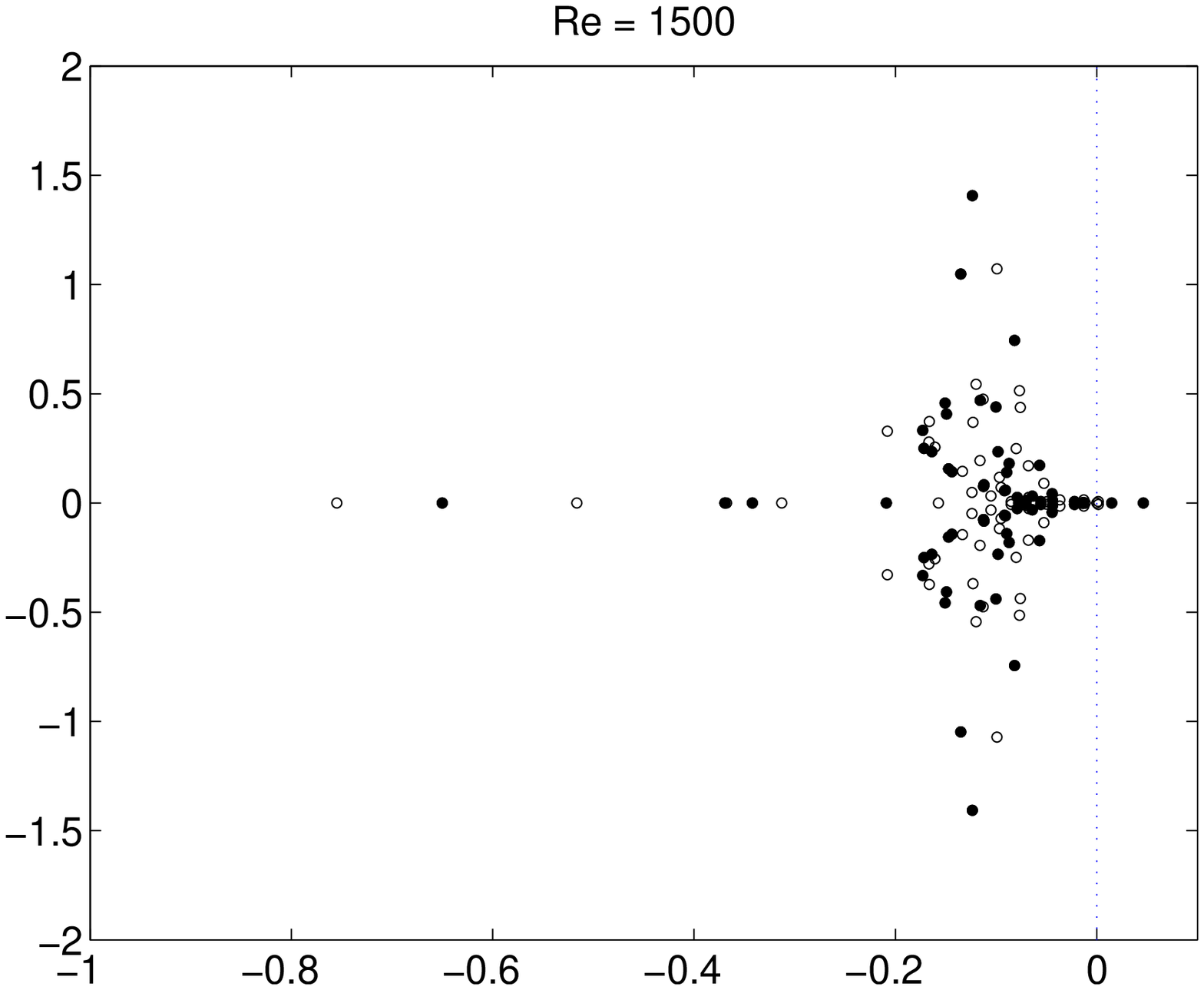}
\end{center}
\caption[xyz]{(a) and (b): Plots of $\mu = \exp(\lambda T)$, where
 $\lambda$ is an eigenvalue of the asymmetric traveling wave and
$T$ is listed in Table \ref{table-1}. The markers are filled in
if the corresponding eigenvectors lie in the shift-reflection invariant
subspace. (c): Plot of the eigenvalues of the asymmetric traveling wave.}
\label{fig-7}
\end{figure}

The approximations $\mu$ and $y$ must be checked for correctness. If
$\mu$ is real, one only has to apply the matrix $\eqref{eqn-5-8}$ to
$Q_k y$ and verify if the resulting vector has the right amplitude and
direction. If $\mu$ is complex, one has to apply the matrix to the
real part of $Q_k y$. In Figures \ref{fig-7} and \ref{fig-8}, we
accept an eigenvalue if the result of applying the matrix has an error
in direction that is less than $1$ degree and the error in amplitude
is less than $1\%$. Most eigenvalues and eigenvectors are much more
accurate than that, and it is reasonable to expect the eigenvalues to
be more accurate than the eigenvectors.

If $x_0$ is the initial velocity field of a traveling wave, its
wavespeeds are given by $c_\theta = -(s_\theta + 2\pi p)/T$ and $c_z =
-(s_z + 2\pi\Lambda q)/T$, where $p$ and $q$ are integers.  The values
of $p$ and $q$ are found by advancing the initial velocity field by 
an amount of time that is not too large, and then translating and
rotating the final velocity field to see which values of $p,q$ imply
the best match to the initial velocity field. In the case of the
asymmetric traveling wave, $c_\theta = s_\theta = 0$ because of
symmetry and care is needed for determining $c_z$ at high $Re$ because
there is very little energy in the streamwise modes with $n \neq 0$.

In the case of traveling waves, there is a delicate numerical point
that arises in passing from a complex eigenvalue $\mu$ of
\eqref{eqn-5-8} to an eigenvalue $\lambda = \log(\mu)/T$ of the
traveling wave.  Figure \ref{fig-7}c shows the $\lambda$s that
correspond to the $\mu$s in Figure \ref{fig-7}a. The imaginary part of
the complex $\log$ is not unique, and to determine it for the
$\lambda$s one has to in effect determine the rate of rotation of the
real part of the eigenvector in the space spanned by the real and
imaginary parts. If the column $c$ is the real part of the eigenvector, the 
matrix-vector product
\begin{equation*}
\exp(c_\theta t\mathcal{T}_1)
\exp(c_z t \mathcal{T}_2) 
\frac{\partial X(t; x_0)}{\partial x_0} c
\end{equation*}
for $t$ not too large will give the correct rate of rotation. To find
that matrix-vector  product, we can again use differences
as in \eqref{eqn-5-6} but there are two mathematically equivalent
ways to do so.
The first way is to use the quotiented difference
\begin{equation}
\frac{\exp(c_\theta t \mathcal{T}_1)
\exp(c_z t\mathcal{T}_2)
X(t; x_0 + \epsilon c)-y_0}{\epsilon},
\label{eqn-5-9}
\end{equation}
where $y_0 = \exp(c_\theta t \mathcal{T}_1 + c_z t\mathcal{T}_2)X(t;x_0)$
is determined using the same direct numerical simulation code 
and the same time step used to
compute $X(t; x_0 + \epsilon c)$,
and the second way is to use 
\begin{equation}
\frac{\exp(c_\theta t \mathcal{T}_1)
\exp(c_z t\mathcal{T}_2)
X(t; x_0 + \epsilon c) - 
x_0
}{\epsilon}.
\label{eqn-5-10}
\end{equation}
We must use
\eqref{eqn-5-9}, although \eqref{eqn-5-10} involves less work.
The numerical errors in using the quotiented difference \eqref{eqn-5-10} will
be intolerably high.

The eigenvalues in Figure \ref{fig-7}a,b  are mostly
inside the unit circle and stable. Most of the eigenvalues of the
matrix \eqref{eqn-5-8} are stable because of the dissipation term in
the Navier-Stokes equation.  For a demonstration of the effect of the
dissipation term, note that the stable eigenvalues for $Re=1500$ are
closer to the circle than those of $Re=15000$, even though the
computation at $Re=15000$ uses a larger $T$ (see Table
\ref{table-1}) which brings the stable eigenvalues closer to the
center. 


Setting up the eigenvalue problem for traveling waves using direct
numerical simulation and the matrix \eqref{eqn-5-8} may seem contrived
because of the need to choose an artificial parameter $T$ 
and the need to use direct numerical simulation. Contrived it may be,
but the contrivance does serve a purpose. Without it we will have a 
spectrum that will look like the one in Figure \ref{fig-7}c, but with
a lot of eigenvalues with very large and negative real parts not
shown there. For a matrix with such a spectrum, the Arnoldi iteration
will not work well because it will be forced to chase the eigenvalues
with large and negative real parts. With matrix \eqref{eqn-5-8}, those
eigenvalues move very close to $0$, and the extremal part of the
spectrum that is approximated well is also the interesting part
of the spectrum for stability considerations.

\section{Spectrum of lower-branch traveling waves as $Re\rightarrow\infty$}
\begin{figure}
\begin{center}
(a)\includegraphics[height=1.5in,width=1.8in]{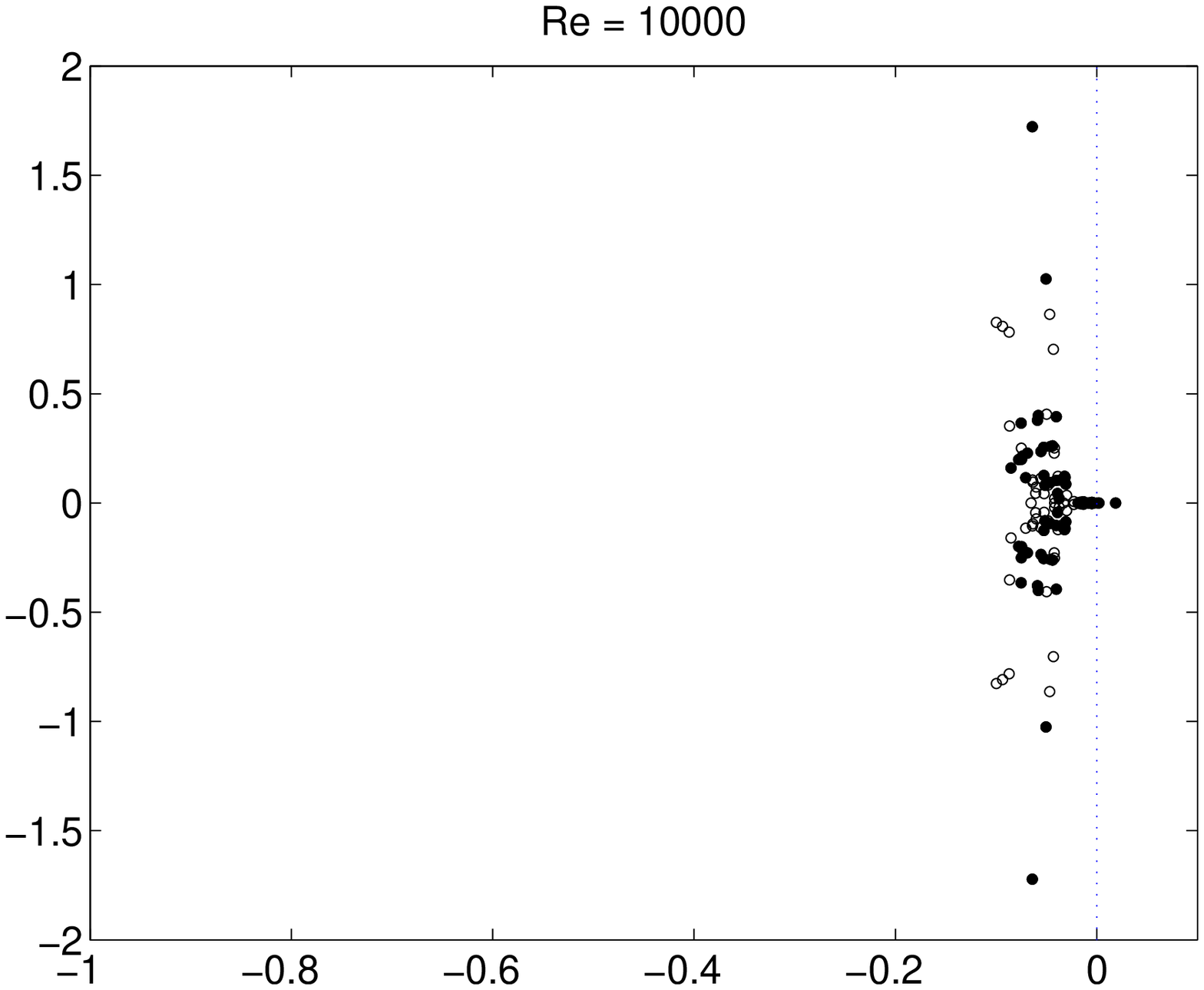}
(b)\includegraphics[height=1.5in,width=1.8in]{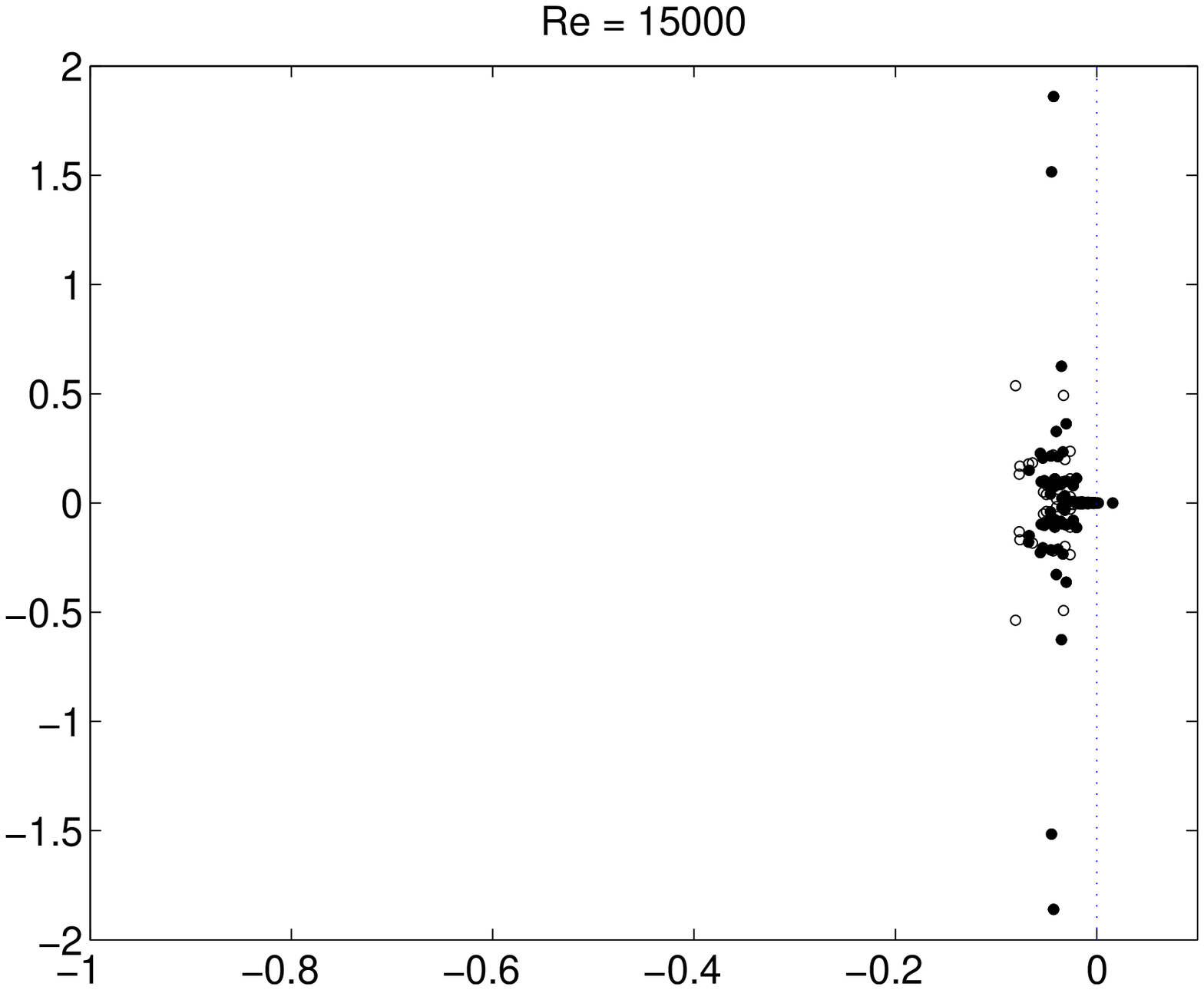}
(c)\includegraphics[height=1.5in,width=1.8in]{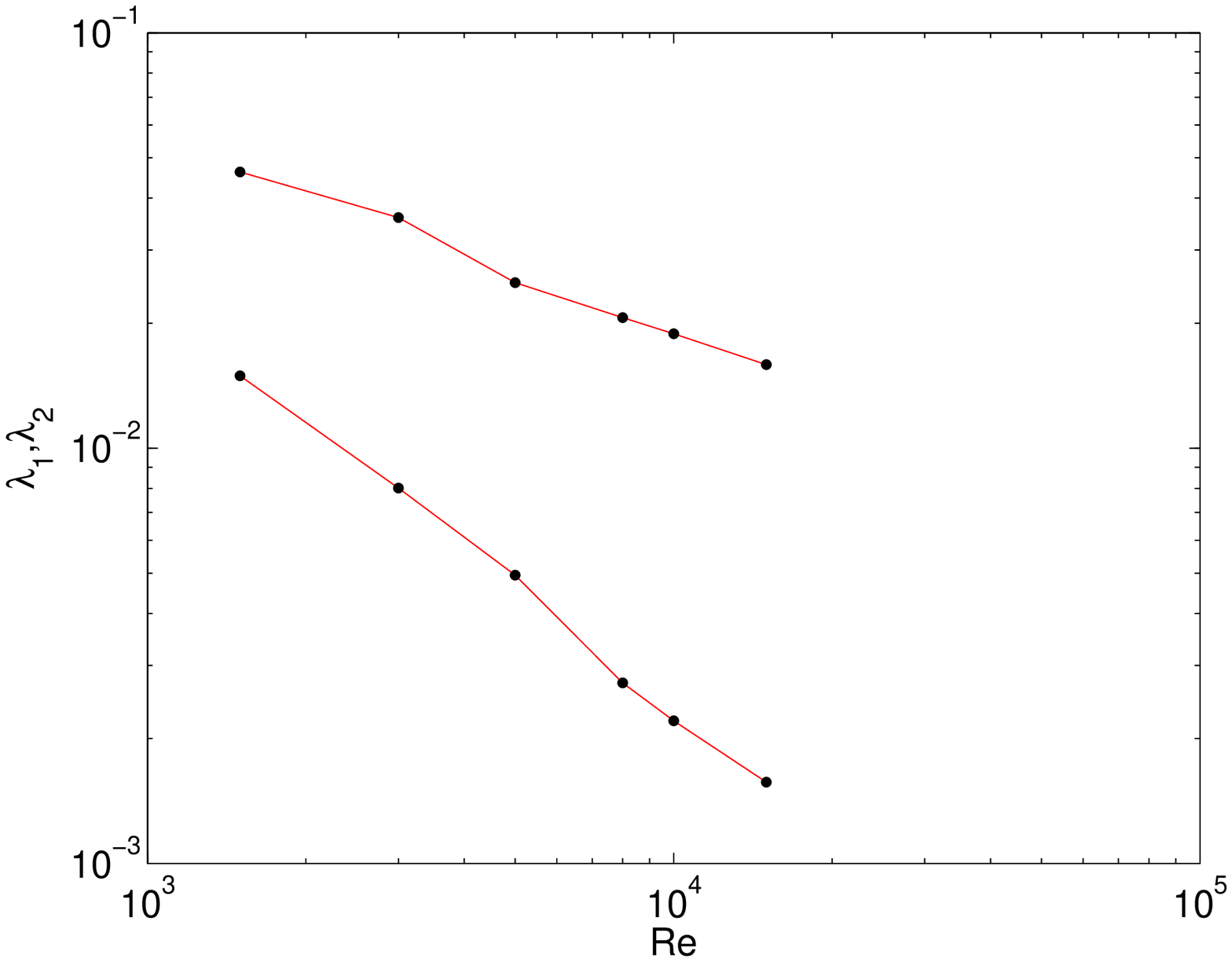}
\end{center}
\caption[xyz]{(a) and (b): 
Plots of eigenvalues of the asymmetric traveling wave.
(c): Scaling of the two unstable eigenvalues in the shift-reflection
invariant subspace as $Re\rightarrow\infty$.}
\label{fig-8}
\end{figure}

The Arnoldi iterations for  traveling waves at various $Re$ were
carried out using $k=150$. For $Re=1500$, $122$ out of $150$ eigenvalues
of $H_k$ turned out to be correct. For $Re=15000$ as well, $122$
out of the $150$ eigenvalues were correct, but the time of integration
was higher with $T=15$.

At $Re=1500$, the asymmetric traveling wave has two real unstable
eigenvalues, whose eigenvectors are invariant under shift-reflection.
Those two eigenvalues persist as $Re\rightarrow \infty$. Surprisingly,
those two eigenvalues approach $0$ as $Re\rightarrow \infty$. 
Figure \ref{fig-8}b shows that the rate of decrease of those eigenvalues
is algebraic. The most unstable eigenvalue approaches $0$ at the
rate $Re^{-0.41}$. The other eigenvalue approaches $0$ at the faster
rate $Re^{-0.87}$. For the symmetric lower-branch solution of 
plane Couette flow, there is just one unstable eigenvalue and that
decreases at the rate $Re^{-0.46}$ or $Re^{-0.48}$ \citep{Viswanath2, WGW}.
Figure \ref{fig-8}a,b shows that the spectrum as a whole approaches the
imaginary axis as $Re$ increases.

In addition to the two real unstable eigenvalues with eigenvectors in
the symmetric subspace, there is an unstable complex pair at $Re=1500$
which can be seen in Figure \ref{fig-7}a. That pair moves  inside
the circle as $Re$ increases. At $Re = 3000$ and $Re=5000$, there is 
a third real and weakly unstable eigenvalue. For $Re\geq 8000$,
there seem to be only two unstable eigenvalues, and both of those
have eigenvectors that are invariant under shift-reflection.

\section{Conclusion}

We have demonstrated the existence of a critical layer in the
$Re\rightarrow\infty$ limit for a family of lower-branch traveling
waves. The theory of \cite{WGW} gives the right formula for the
critical curve. The scaling of the size of the critical region for
$\abs{u_1}$ is in excellent agreement with their theory. Further development
of the asymptotic theory appears necessary to explain the scaling of
the size of the critical regions for $\abs{w_1}$ and $\zeta_0$. 
Comparison with a family of lower-branch equilibrium solutions of
plane Couette flow suggests that the formation of the critical layer
and many of its properties could be universal to all lower-branch solutions
of shear flows as $Re\rightarrow\infty$.

Certain parts of puffs, which are structures observed in transitional
pipe flow, are characterized by streaks and rolls \citep{HD,WillisK2}. We
have suggested that the critical surface of a puff could be helpful in
visualizing its structure. In particular, the arrangement of rolls and
streaks could be correlated with the shape of the critical surface.

In Section 5, we have given a detailed account of the GMRES-hookstep
iteration for computing relative periodic solutions, traveling waves,
periodic solutions, and equilibria for shear flows. Our account
emphasizes the implementation aspects of GMRES-hookstep and of the
Arnoldi iteration, which is used for finding eigenvalues. Together
with the derivation of the Newton equations \citep{Viswanath1}, this
account is sufficiently detailed to enable implementation of these
iterations. 

\begin{acknowledgements}
The author thanks the mathematics department of the Indian Institute
of Science, Bangalore, for its hospitality and support.
The author thanks F. Waleffe and J.F. Gibson for helpful discussions,
and W.R. Morrow for catching a bad typo.
This work was partially supported by NSF grants DMS-0407110 and
DMS-0715510.
\end{acknowledgements}


\bibliographystyle{agsm}
\bibliography{references}

\label{lastpage}

\end{document}